\documentclass[twocolumn]{aastex63}

\hypersetup{linkcolor=blue,citecolor=blue,filecolor=cyan,urlcolor=magenta}

\shorttitle{Distances to radio spurs I and IV}
\shortauthors{Panopoulou et al.}

\graphicspath{{./}{figures/}}

\usepackage{amsmath}
\usepackage{comment}
\usepackage{lineno}

\begin{document}

\title{Revisiting the distance to radio Loops I and IV using Gaia and radio/optical polarization data}

\correspondingauthor{G. V. Panopoulou}
\email{panopg@caltech.edu}

\author[0000-0001-7482-5759]{G. V. Panopoulou}
\altaffiliation{Hubble Fellow}
\affiliation{Cahill Center for Astronomy and Astrophysics, California Institute of Technology, MC 350-17, Pasadena, CA 91125, USA}

\author[0000-0002-0045-442X]{C. Dickinson}
\affiliation{Jodrell Bank Centre for Astrophysics, Alan Turing Building, Department of Physics and Astronomy, School of Natural Sciences, \\
The University of Manchester, Oxford Road, Manchester, M13 9PL, Manchester, U.K.}

\author[0000-0001-9152-961X]{A. C. S. Readhead}
\affiliation{Cahill Center for Astronomy and Astrophysics, California Institute of Technology, MC 350-17, Pasadena, CA 91125, USA}
\affiliation{Owens Valley Radio Observatory, California Institute of Technology, Pasadena, CA 91125, USA}

\author[0000-0001-5213-6231]{T. J. Pearson}
\affiliation{Cahill Center for Astronomy and Astrophysics, California Institute of Technology, MC 350-17, Pasadena, CA 91125, USA}
\affiliation{Owens Valley Radio Observatory, California Institute of Technology, Pasadena, CA 91125, USA}

\author[0000-0003-3412-2586]{M. W. Peel}
\affiliation{Instituto de Astrof\'{i}sica de Canarias, E-38205 La Laguna, Tenerife, Spain}
\affiliation{Departamento de Astrof\'{i}sica, Universidad de La Laguna (ULL), E-38206 La Laguna, Tenerife, Spain}

\begin{abstract}
Galactic synchrotron emission exhibits large-angular-scale features known as radio spurs and loops.
Determining the physical size of these structures is important for understanding the local interstellar structure and for modeling the Galactic magnetic field. However, the distance to these structures is either under debate or entirely unknown. We revisit a classical method of finding the location of radio spurs by comparing optical polarization angles with those of synchrotron emission as a function of distance. We consider three tracers of the magnetic field: stellar polarization, polarized synchrotron radio emission, and polarized thermal dust emission. 
We employ archival measurements of optical starlight polarization and \textit{Gaia} distances, and construct a new map of polarized synchrotron emission from WMAP and \textit{Planck} data. We confirm that synchrotron, dust emission, and stellar polarization angles all show a statistically significant alignment at high Galactic latitude. We obtain distance limits to three regions towards Loop I of 112$\pm$17\,pc, 135$\pm$20\,pc, and $<105$\,pc. Our results strongly suggest that the polarized synchrotron emission towards the North Polar Spur at $b > 30^\circ$ is local. This is consistent with the conclusions of earlier work based on stellar polarization and extinction, but in stark contrast with the Galactic center origin recently revisited on the basis of X-ray data. We also obtain a distance measurement towards part of Loop\,IV (180$\pm$15\,pc) and find evidence that its synchrotron emission arises from chance overlap of structures located at different distances. Future optical polarization surveys will allow the expansion of this analysis to other radio spurs.
\end{abstract}


\section{Introduction} \label{sec:intro}

The low frequency ($\lesssim 70$\,GHz) radio sky at high Galactic latitudes is dominated by bright, well-known, large-scale features \citep{Berkhuijsen1971}. They have steep radio spectra and are highly polarized, indicative of synchrotron radiation \citep[for determinations of the spectral indices see][]{Vidal2015,Planck2015_XXV}. Some of these features appear to trace full or partial circles on the sky, and are referred to as `loops' or `arcs'. The term `spurs', introduced by \cite{Berkhuijsen1971}, is used to describe continuous regions of bright synchrotron intensity that do not necessarily trace circles. As early as their discovery, it was recognized that the structures are likely in the Galactic neighborhood of the Sun, which explains their large (tens of degrees) angular sizes. The common assumption on the origin of these structures is that they are the remnants of one, or a series of supernova explosions \citep[for a review see][]{Salter1983}. 

Understanding these large-scale objects is important for a number of reasons. They are physical structures that are likely nearby and are therefore an important component for understanding the local interstellar environment. If indeed they are old supernova remnants, they trace historical star-formation events that occurred in the vicinity of the Sun (within the nearest kiloparsec). They also allow the detailed study of such objects before they fade into the general background due to radiative cooling \citep[e.g.,][and references therein]{Sarbadhicary2017}. Measuring the large-scale magnetic field (henceforth, $B$-field) of our Galaxy is a major endeavour that is important for understanding the Galactic dynamo and the structure of the Galactic disk \citep{PIP_XLII}. Whether nearby or distant, radio spurs and their local $B$-fields can cause confusion for studies attempting to characterize the large-scale Galactic $B$-field \citep[e.g.,][]{Jaffe2019}. Furthermore, modeling the emission of these structures could be important for understanding and removing polarized foreground emission from sensitive Cosmic Microwave Background (CMB) surveys \citep[e.g.,][]{Dunkley2009a,Remazeilles2018}.

The physical locations of most of these features are either debated between the proponents of rival interpretations or entirely unknown. A prominent example is the North Polar Spur (NPS), which is the brightest part of the larger Loop\,I, rising
vertically above the Galactic plane at a longitude of $\sim 30^\circ$ and having a diameter of $\sim 100^{\circ}$ on the sky \citep{Large1962,Berkhuijsen1971}. Soon after its discovery, \cite{Bingham1967} found that the electric vector position angles (polarization angles) of a small number ($\approx 10$) of stars were aligned with the $B$-field direction traced by synchrotron polarization, thus suggesting a distance of $\sim$ 100\,pc to the high-latitude part of Loop\,I. Subsequent studies of the optical polarization have confirmed this conclusion \citep{Spoelstra1972,ellisaxon1978,leroy1999,Santos2011,Berdyugin2014}, and are in agreement with the distance to the neutral gas likely associated with the NPS \citep{PuspitariniLallement2012,Das2020}. In apparent contradiction with these determinations, an alternative model has been proposed for Loop I, in which the structure is located at the Galactic center and forms part of a bipolar hyper-bubble extending out to the Galactic halo \citep{Sofue1977, Sofue2000}. In partial support of this model, the absorption of X-rays associated with the NPS necessitates a much larger distance than 100\,pc \citep[300\,pc--4\,kpc;][]{Sofue2015,Lallement2016}. Recently, several studies of X-ray data have argued in favor of the NPS being at the Galactic centre at 7.8\,kpc \citep[]{Akita2018,Predehl2020,Kataoka2021}, possibly related to the \textit{Fermi} $\gamma$-ray bubbles \citep{Dobler2010}. The debate has been reviewed several times in the past few years \citep[][]{Planck2015_XXV,Dickinson2018,Kataoka2018}.

With the influx of new data on the Galactic $B$-field we can hope to gain better insights on these interesting radio features and their relation to the $B$-field on large scales \citep[e.g.,][]{PIP_XLII}. New maps of polarized synchrotron and thermal dust emission \citep{Bennett2013,Planck2018_I} and stellar polarization surveys \citep[e.g.,][]{Berdyugin2014,Clemens2020} can provide measurements of the $B$-field geometry with unprecedented detail (see Section\,\ref{sec:data}). In conjunction with the $B$-field tracers, we can use stellar distance measurements from \textit{Gaia} \citep{Gaiamission2016}.

The basis of the method for placing distance limits to synchrotron spurs is to search for alignment between stellar and synchrotron polarization angles. Alignment can be achieved under certain circumstances. First, the synchrotron polarization along a line of sight is dominated by the emission local to the spur (i.e., there is no significant confusion from background/foreground emission), and Faraday rotation is negligible (the latter is true far from the Galactic plane at frequencies higher than 10\,GHz). If this is the case, the intrinsic synchrotron polarization angles at emission trace the magnetic field geometry local to the spur. For the brightest spurs (e.g., Loop\,I) this is likely the case, as the polarization angles are closely aligned with the axis of the structure \citep{Vidal2015}. Second, if the synchrotron spurs are indeed the result of $B$-fields that have been compressed along the surface of an evacuated cavity \citep[by a supernova explosion e.g.,][]{Ferriere1991}, then it is likely that dust will also remain on the surface. The dust grains induce polarization to light from background stars. Since starlight polarization is an integrated effect of dust along the line-of-sight to the star; the dust that is colocated with the synchrotron emission will give rise to stellar polarization that traces the same $B$-field probed by the synchrotron emission. Thus a significant correlation between stellar and synchrotron polarization angles allows us to infer that the synchrotron emission is closer than the stars that show the correlation.

By using many stars in a given region of sky, and looking for alignment of their polarization angles with the $B$-field orientation inferred from synchrotron/dust emission data, we can statistically infer a distance to the synchrotron spurs \citep[as originally done by][]{Bingham1967,Spoelstra1972}. Of course, the conditions for alignment may not be met. An alternative hypothesis can also be tested: if the $B$-field orientations are not aligned in a given region, then the synchrotron and thermal dust emitting ISM must be at different distances. 

In this paper we revisit the comparison between the polarization of starlight and synchrotron emission, aiming to improve the existing distance determination towards synchrotron loops and spurs. Our analysis brings together a recently compiled catalog of stellar polarization measurements, as well as improved synchrotron data obtained by combining WMAP and {\it Planck} low frequency maps. We make use of three additional datasets: (a) accurate stellar distances from the \textit{Gaia} mission, (b) the latest \textit{Planck} polarized dust emission map and (c) 3D maps of the stellar extinction in the Solar neighborhood. We introduce the data and pre-processing in Section \ref{sec:data}. The comparison between the three different polarization datasets is presented in Section \ref{sec:results}. We determine the distance to two structures: Loop\,I and Loop\,IV in Sections  \ref{sec:results_loopI}, \ref{sec:results_loopIV} and \ref{sec:limits}.  We discuss our findings in Section \ref{sec:discussion} and conclude in Section \ref{sec:conclusions}.

\begin{figure*}[ht]
\centering
\includegraphics[scale=1]{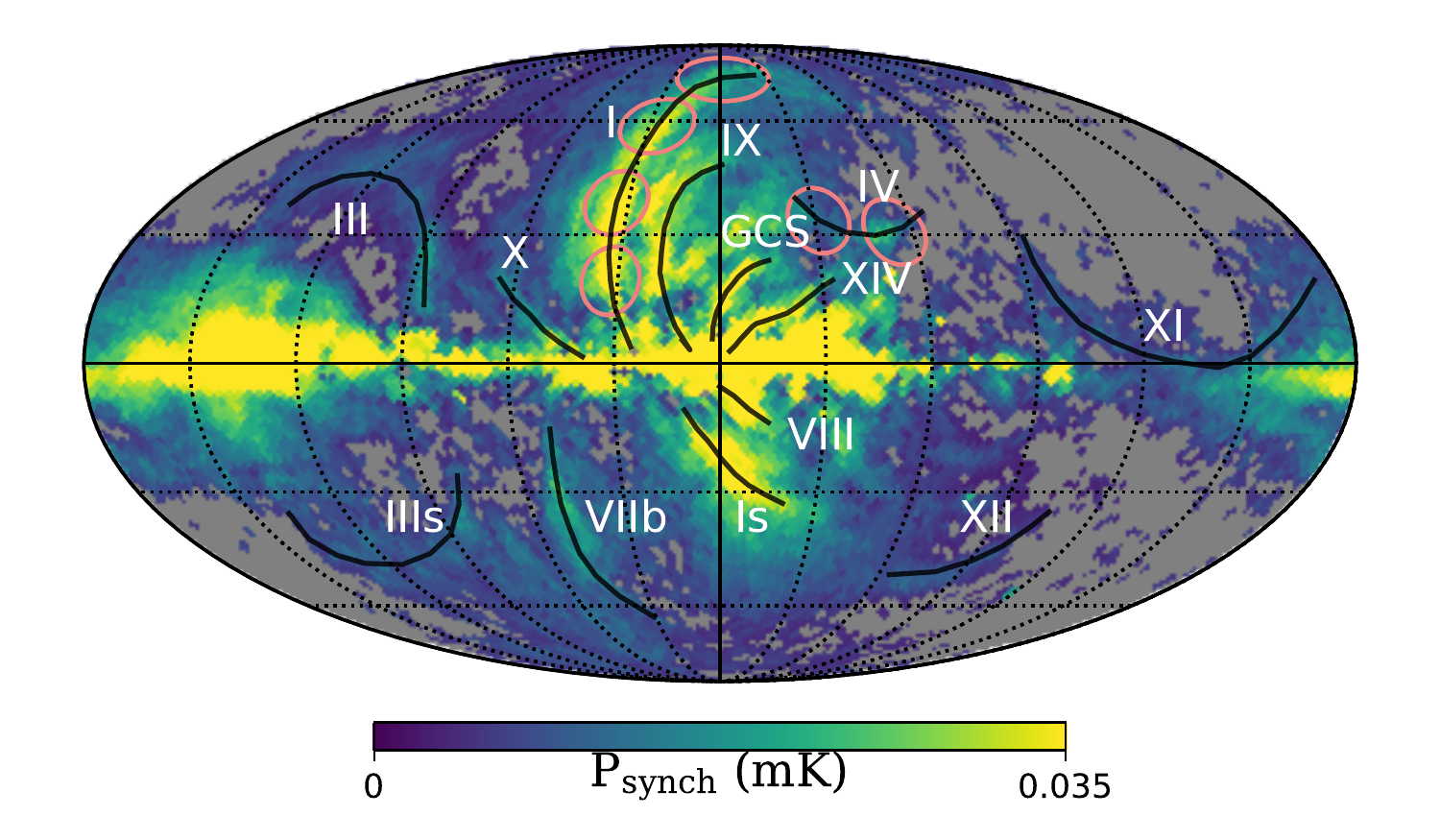}
\caption{All-sky map of polarized intensity of synchrotron emission at 28.4\,GHz obtained by combining  WMAP/{\it Planck} LFI data at $N_{\rm side}=32$ (see Section~\ref{sec:data}). The main loops/spurs of bright polarized emission from \protect\cite{Vidal2015} are shown with black lines. Pixels with $\rm S/N< 4$ are masked (grey). The map is in Mollweide projection, centered at $l,\,b = (0^\circ,0^\circ)$, with grid lines spaced by 30$^\circ$. The colorscale is linear, with a threshold at 0.035 mK to highlight the fainter portions of the map. Pink circles mark the regions analysed in Section \ref{sec:results}.}
\label{fig:loops}
\end{figure*}

\section{data \& pre-processing} \label{sec:style}
\label{sec:data}

\subsection{Polarized synchrotron emission}
\label{sec:synchrotron}

Synchrotron radiation is produced by relativistic electrons spiralling in a magnetic field. If the field is regular and ordered, the polarization fraction can be as high as $\approx 75$ per cent \citep{Rybicki_book}. The measured polarization angle traces the magnetic field orientation projected on the sky. If there is no depolarization (e.g., along the line-of-sight) then this provides a direct way of measuring the orientation of the $B$-field.

At frequencies of a few GHz and higher, the radiation is optically thin and has a steep spectrum falling with frequency \citep[with a brightness temperature spectral index $\beta$, of $\approx -3$;][]{Bennett2013,Planck2015_XXV,Fuskeland2021}. Thus low frequencies ($\lesssim 1$\,GHz) would provide a very bright signal. However, at frequencies below a few GHz, the polarization angle is rotated under the effect of Faraday Rotation \citep[e.g.,][]{Wolleben2006,Carretti2019,Hutschenreuter2020}. This can depolarize the observed signal and means that the angle is no longer perpendicular to the $B$-field.

Fortunately, WMAP \citep{Bennett2013} and {\it Planck} \citep{Planck2018_I} have mapped the entire sky at frequencies above 20\,GHz in intensity and polarization. At these frequencies, Faraday Rotation is negligible ($\ll1^{\circ}$ rotation) except for regions near the Galactic Centre \citep{Vidal2015}. The main limitation with these data is the low signal-to-noise ratio (S/N), particularly in areas of the sky where there is little polarized synchrotron emission. The lowest frequency channels of these two surveys provide the highest S/N; WMAP K-band at 22.8\,GHz and {\it Planck} LFI at 28.4\,GHz.

To increase the S/N, we can smooth and degrade the maps to a lower angular resolution. However, even after smoothing there remain large areas of the sky where the S/N is very low. To increase the S/N further, we combine the three lowest channels of WMAP (22.8, 33.0\, and 40.7\,GHz) with {\it Planck} data at 28.4 and 44.1\,GHz \citep[for more details see Peel et al., in prep., which follows a similar technique to][]{Planck2015_XXV}. We use the Stokes $Q,U$ maps, which are provided in HEALPix format \citep{Gorski2005} on the Planck Legacy Archive\footnote{\url{https://pla.esac.esa.int/pla/}}. We first smooth the $Q,U$ maps to a common $1^{\circ}$ angular resolution, using the published beam window functions, and degrade to a given HEALPix $N_{\rm side}$. We then include color corrections and extrapolate each pixel to a common frequency of 28.4\,GHz assuming a spectral index $\beta=-3.1$. This is the average value measured at these frequencies and does not appear to vary much across the sky \citep{Bennett2013,Fuskeland2021}, except for possibly a few areas near the Galactic plane \citep{PIP_IX}. We then make the weighted average of these data \citep[see also][]{Planck2015_XXV}, based on the noise covariance matrices computed using Monte Carlo noise realizations. The noise realizations make use of the published $QQ$, $QU$ and $UU$ variance maps, with each realization smoothed to 1$^\circ$ before combining to calculate the smoothed covariance matrices at a given $N_{\rm side}$ \citep[similarly to][see Peel et al., in prep.]{PIP_XIX}. The majority of the weight comes from WMAP 22.8\,GHz, with some regions dominated by {\it Planck} 28.4\,GHz. Nevertheless, there is an improvement in the S/N of $\approx 2$ on average.

Fig.\,\ref{fig:loops} shows the polarized intensity at 28.4\,GHz at a resolution parameter of $N_{\rm side}=32$ (pixels $\approx 1^{\circ}\!.8$ on a side). Pixels with $\rm S/N<4$, corresponding to an angle uncertainty of $\sigma_\phi > 7.\!^{\circ}1$, have been masked and are not used (see Section \ref{sec:angdiffs}). The bright loops and spurs seen in the map are detected at high significance.

\subsection{Polarized thermal dust emission}
\label{sec:thermaldust}

Thermal dust emission from cold dust ($T \sim 10$--100\,K) dominates the continuum at frequencies higher than $\sim300$\,GHz \citep{PIP_XVII}. Polarized emission arises from non-spherical dust grains aligning their short axis with the $B$-field. Thus, as in synchrotron emission, the polarization angle is perpendicular to the $B$-field. {\it Planck} has mapped the polarized thermal dust emission at frequencies $100$--353\,GHz with the HFI instrument \citep{PIP_XIX}. The highest frequency (353\,GHz) has the highest signal-to-noise ratio due to the steeply rising spectrum of thermal dust emission. 

For our analysis we use the CIB-filtered common resolution map at 353\,GHz  \citep{Planck2018_XII,Planck2018_IV}, which provides the maximum signal-to-noise ratio. We mask pixels with $\rm S/N<4$ and apply the same degradation as for the synchrotron map to a common $N_{\rm side}$.

\subsection{Starlight polarization catalog}
\label{sec:stars}

We use a new compilation of optical polarization data (Panopoulou et al., in prep.). The catalog contains $\sim$ 35,000 stars with measured polarization. The data cover a wide range of environments, from molecular clouds to the diffuse, high Galactic-latitude ISM. Quantities in the catalog include: stellar polarization fraction, $p$, polarization angle in Galactic coordinates, $\theta^*$, and associated uncertainties. The catalog also provides a cross-match with \textit{Gaia} DR2 \citep{gaiadr2}. Distances are obtained from \cite{Bailer-Jones}, and include a point estimate as well as 68\% confidence intervals. For stars that are not included in \textit{Gaia} DR2, or do not have distance estimates in \citet{Bailer-Jones}, distances were obtained by inverting the \textit{Hipparcos} parallax, where available.
We discard any entries in the catalog where the polarization angle uncertainty, $\sigma_\theta^*$, is not provided. We convert all polarization angles to the range $[-90^\circ,90^\circ)$. If there are multiple measurements of the same star (matching coordinates to within a radius of 1-arcsecond), we use the measurement with smallest uncertainty in polarization angle.

For comparison with the polarization angles in emission, we bin the data into HEALPix maps of ${N_{\rm side}} = 32$ ($1.\!{^\circ}8$).
Within each HEALPix pixel, we calculate the weighted mean polarization angle, $\theta$, of the $N$ stars in the pixel:

\begin{equation}
    \theta = \frac{1}{2} \, \arctan \left[ \frac{1}{W}\sum_{i=1}^{N} w_i \sin(2\theta_i^*), \frac{1}{W}\sum_{i=1}^{N} w_i \cos(2\theta_i^*)\right],
    \label{eqn:circmean}
\end{equation}
as appropriate for circular quantities \citep[e.g.,][]{Fisher1995}, where $\rm arctan$ is the two-argument arctangent function. The weights are: $w_i = (\sigma^*_{\theta_i})^{-2}$
and $W$ is the sum of the weights, $\sum_{i=1}^N w_i$.
For stars with very large uncertainties, i.e. $\sigma_{\theta^*} > 45^\circ$, we assign a weight of 0. To avoid assigning excessively large weights to stars with $\sigma_{\theta^*} < 1^\circ$, we set a minimum uncertainty of $1^\circ$ to these stars. This limit also reflects a systematic uncertainty in the absolute calibration of the polarization angle, which arises from the intrinsic variability of calibrator stars \citep[][]{Ramaprakash2019}.

Throughout the text, we use the symbol $\theta^*$ to denote the polarization angle of an individual star. We use the symbol $\theta$ to denote the polarization angle of stars averaged within a sky pixel as described above. All angles are in Galactic coordinates, measured according to the IAU convention from Galactic North through East.

\subsection{Angle differences}
\label{sec:angdiffs}

Our analysis makes use of polarization data from three different tracers of the magnetic field: dust emission, synchrotron emission, and starlight absorption. While starlight polarization angles are parallel to the magnetic field, those of synchrotron and dust emission are perpendicular to it. We rotate the polarization angles of the synchrotron, $\phi_{\rm synch}$, and dust emission, $\chi_{\rm dust}$, by $90^\circ$ for the entirety of the analysis: 

\begin{equation}
    \phi_{\rm synch} = \frac{1}{2}  \arctan(-U_{\rm synch},Q_{\rm synch}) - 90^\circ
\end{equation}
\begin{equation}
    \chi_{\rm dust} = \frac{1}{2}  \arctan(-U_{\rm dust},Q_{\rm dust}) - 90^\circ
\end{equation}
where $\rm arctan$ is the two-argument arctangent function, while $U_{\rm synch},Q_{\rm synch}$ and $U_{\rm dust},Q_{\rm dust}$ are the Stokes parameters of the synchrotron and dust emission maps (Sections \ref{sec:synchrotron}, \ref{sec:thermaldust}), respectively. 
We have multiplied the Stokes parameter $U$ by $-1$ to convert from the COSMO convention to IAU \citep[e.g.,][]{Planck2018_XII}. 
We transform each angle $\psi'$ by:
\begin{equation}
    \psi = (\psi' + 90^\circ) {\rm mod} (180^\circ) - 90^\circ,
\end{equation}
so that the final angle $\psi$ is defined in the range $[-90^\circ, 90^\circ)$. 
We make use of the following angle differences between the three magnetic field tracers: $\theta^* - \phi_{\rm synch}, \, \theta^* - \chi_{\rm dust},$ and $\chi_{\rm dust} - \phi_{\rm synch}$.


We wish to study regions where we can be confident of the alignment between different magnetic field tracers. For this reason, we
apply a S/N threshold on the polarized intensity of synchrotron emission and dust emission of $\rm S/N > 4$, which corresponds to an uncertainty in the polarization angle of $7^\circ$. Thus we mask any regions in our maps with S/N lower than this threshold.

We aim to use the observed alignment between the three $B$-field tracers to constrain our estimates of the distance to a synchrotron spur. However, if the synchrotron emission has significant contributions from background or foreground components along the line-of-sight, then the observed synchrotron polarization angle may not represent the $B$-field orientation local to the spur, depending on the relative brightness of each component. For a foreground or background component whose polarized intensity is 60\% that of the synchrotron spur, the maximum angle difference between the observed polarization angle and that of the $B$-field at the spur's location is 20$^\circ$. We exclude potentially problematic situations by requiring that the observed offset between the synchrotron $\phi_{\rm synch}$ and any other tracer be $<20^\circ$. Furthermore, our analysis is focused on the brightest polarized radio emission from the loops/spurs, thus reducing the possibility of bias due to foreground or background emission. Nevertheless, chance alignments can still occur; for a random distribution of angle differences, there is a 22\% probability that any one measurement is within  $<20^\circ$. However, since we are dealing with multiple (tens of) stellar measurements within a given region, chance alignments are much less likely (of order a few \%). 

To quantify the significance of the alignment between two angles, we use the Projected Rayleigh Statistic \citep[PRS,][]{Jow2018}, which is specifically designed for dealing with orientations  rather than circular quantities. The PRS is an optimal statistic for testing the hypothesis that a distribution of angle differences is consistent with alignment. We use the definition of the PRS that takes into account measurement uncertainties:

\begin{equation}
{\rm PRS} = \frac{1}{\sqrt{\sum_i^N w^2_i/2 }}\sum_i^N w_i \cos{2\Delta\psi_i},
\end{equation}
where $\Delta\psi_i$ is the difference between two polarization angle measurements (e.g., $\theta^*-\phi_{\rm synch}$) and $w_i$ is the weight as defined for equation \ref{eqn:circmean}. 
The derived PRS will be compared to the PRS of a uniform distribution of angle differences in Section \ref{sec:limits}. 

\subsection{Extinction maps}

We use the \citet{Leike2020} map of 3D dust extinction to measure the distribution of interstellar dust as a function of distance. We query the map via the \texttt{dustmaps} Python package \citep{Green2018} and obtain the mean value of \textit{Gaia} $\rm G$ band optical depth per parsec, $\tau_{\rm G}{\rm/pc}$, in each voxel along a line of sight. We convert the native units of the dataset, $\tau_{\rm G}{\rm/pc}$, to $\rm G$-band extinction using the standard relation between optical depth and extinction:

\begin{equation}
A_{\rm G} = 1.086 \,\, \tau_{\rm G}.
\label{eqn:ag_tau}
\end{equation}
The line-of-sight resolution of the map is 1\,pc. We obtain the cumulative extinction out to a certain distance along the line-of-sight by summing the optical depth per parsec over voxels out to that distance (and multiplying by the factor in equation \ref{eqn:ag_tau}).

To obtain an estimate of the total extinction along the line-of-sight, we use the map presented in \citet{planckxxix}\footnote{The filename of the \textit{Planck} total extinction map is:  COM\_CompMap\_Dust-DL07-AvMaps\_2048\_R2.00.fits} which was created by fitting the dust model of \citet{DraineLi2007} to dust emission and renormalized to match quasar extinctions. This map is given in units of magnitudes in the $\rm V$ band, $A_{\rm V}$.

We use the \citet{Fitzpatrick2019} extinction curve to convert the {\it Planck} $A_{\rm V}$ to G-band extinction, $A_{\rm G}$. The pre-launch \textit{Gaia} G-band central wavelength is 0.673\,$\mu $m \citep{Jordi2010}. We therefore take the inverse monochromatic $\rm G$-band wavelength to be $1.5\,\mu$m$^{-1}$, which corresponds to a value of $ E(G-V)/E(B-V) = -0.63$ from table 3 of \citet{Fitzpatrick2019}. We then use the equation 19 in their appendix B to obtain: 

\begin{equation}
    A_{\rm G} = A_{\rm V} \left(\frac{-0.63}{R_{\rm V}}+1\right)=0.796 A_{\rm V},
\end{equation}
for the standard value of $R_{\rm V} = 3.1$.

\begin{figure*}[ht]
\centering
\includegraphics[]{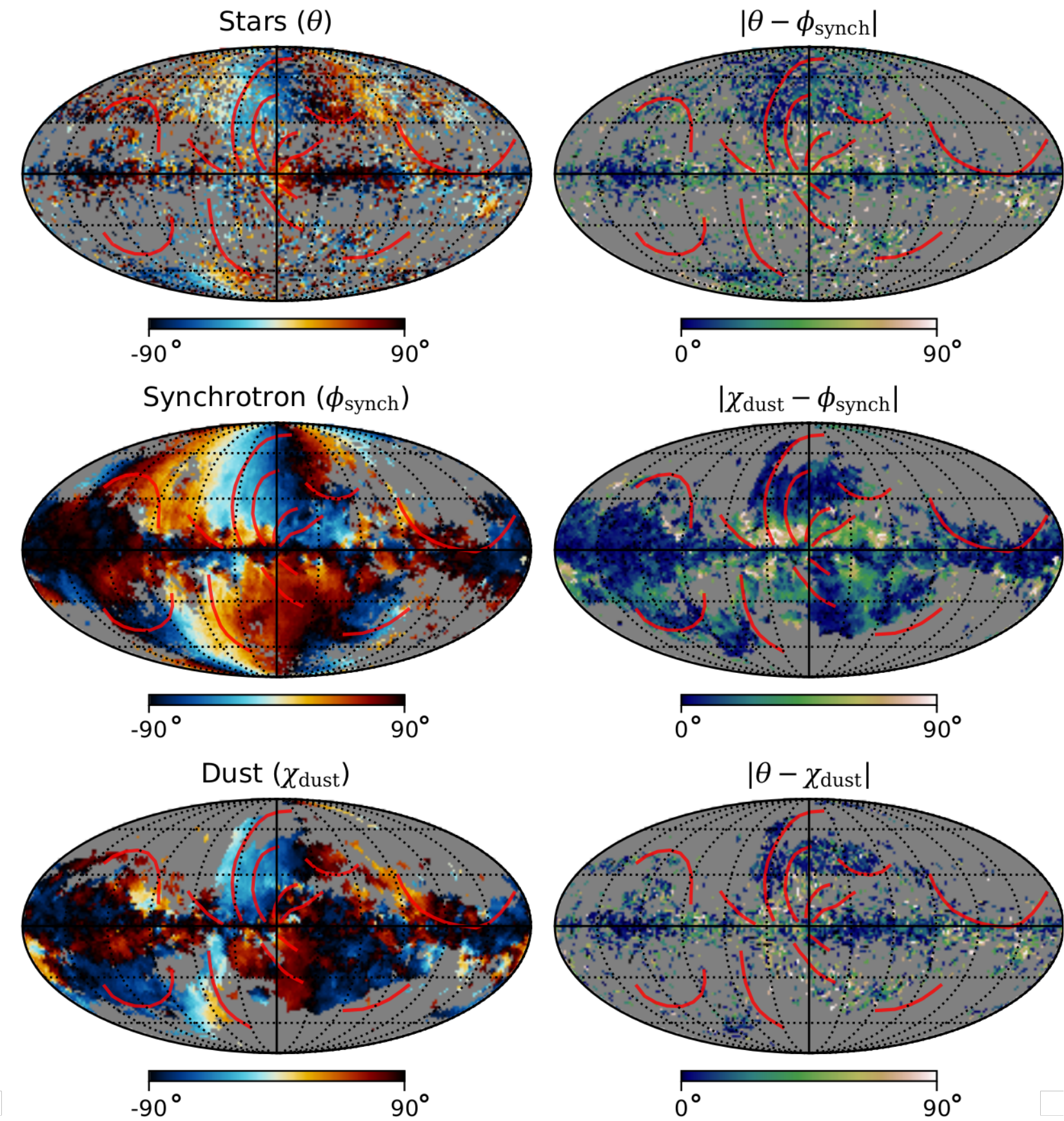}
\caption{Maps of all three $B$-field angle tracers ({\it left} panels) and their pairwise absolute angle differences ({\it right} panels) in degrees. Synchrotron spurs are outlined in bright red (see also Fig.\,\ref{fig:loops}). {\it Left panels}: stellar polarization angle $\theta$ ({\it top}), synchrotron polarization angle, $\phi_{\rm synch}$ ({\it middle}) and  dust emission polarization angle, $\chi_{\rm dust}$ ({\it bottom}). Angles are defined to follow the $B$-field orientation measured in Galactic coordinates from the Galactic North and increasing towards larger Galactic longitude, according to the IAU convention (see Section\,\ref{sec:data}). {\it Right panels}: absolute angle difference between maps of $\theta$ and $\phi_{\rm synch}$ ({\it top}), $\chi_{\rm dust}$ and $\phi_{\rm synch}$ ({\it middle}), and $\theta$ and $\chi_{\rm dust}$ ({\it bottom}). All maps are in Mollweide projection, in Galactic coordinates, centered on $(l, b) = (0^\circ, 0^\circ)$. Grey pixels indicate no data or low S/N ($<4$). Dotted grid lines are spaced by $30^\circ$.}
\label{fig:maps}
\end{figure*}

\section{Results}
\label{sec:results}

We begin in Section\,\ref{sec:results_intro} by briefly discussing the comparison between the three tracers of the $B$-field on large scales. We then discuss our choice for studying two individual loops (I and IV) in more detail, whose results will be presented in the following sections.

\subsection{Large-scale overview and choice of loops}
\label{sec:results_intro}

Figure\,\ref{fig:maps} presents the all-sky maps of the polarization angle of the synchrotron emission, dust emission, and stars as well as the angle differences between all three pairs of tracers at $N_{\rm side}=32$.  Gray pixels denote areas that are masked from the analysis, as mentioned in Section \ref{sec:angdiffs}, either because of low S/N (S/N$<4$) or because they did not contain any stellar polarization measurement. 

As shown in Fig.\,\ref{fig:maps}, there are large areas of sky with no data, even at this relatively low pixel resolution ($N_{\rm side}=32$ or $1.\!{^\circ}8$). The synchrotron map has the best coverage, while the dust emission and starlight maps are more limited. 
Nevertheless, even with limited coverage, it is clear from Fig.\,\ref{fig:maps} that there is a large-scale coherency both in the Galactic plane and at high latitudes in all three tracers. As pointed out before \citep[e.g.,][]{Wolleben2006,Sun2015,PIP_XIX}, this points to a coherent and ordered magnetic field in general, which is also supported by observed high polarization fractions in the synchrotron emission \citep[][]{Vidal2015,Planck2015_XXV}. More importantly, the angle difference maps (right-hand side of Fig.\,\ref{fig:maps}) show remarkable agreement between the three tracers of magnetic field orientation. The overall alignment between these sets of tracers in the plane and at high latitude was already noted for polarization angles of starlight and dust emission \citep{PIP_XXI,Planck2018_XII} while the correlation between synchrotron and dust emission was noted in \citet{PIP_XIX}, particularly in the region around Loop I. 

We note that the $\theta$ map includes only stars out to 1\,kpc. The angle differences between stellar and synchrotron polarization ($|\theta - \phi_{\rm synch}|$) indicate alignment for a substantial fraction of the high latitude sky. We find that 47\% of the pixels at $|b| > 30^\circ$ have $|\theta - \phi_{\rm synch}| < 20^\circ$ (our defined threshold for alignment, see Section\,\ref{sec:angdiffs}). Such widespread alignment was also seen with fewer stellar measurements when comparing to WMAP data \citep{Page2007}. This alignment places an upper limit on the synchrotron emission: in these regions the high latitude polarized synchrotron emission must be dominated by the nearest 1\,kpc. With wider and deeper coverage of stellar surveys we will be able to place more stringent limits on the fraction of emission that is contributed beyond this distance as a function of latitude.

\begin{figure*}[ht]
\centering
\includegraphics[scale=0.74]{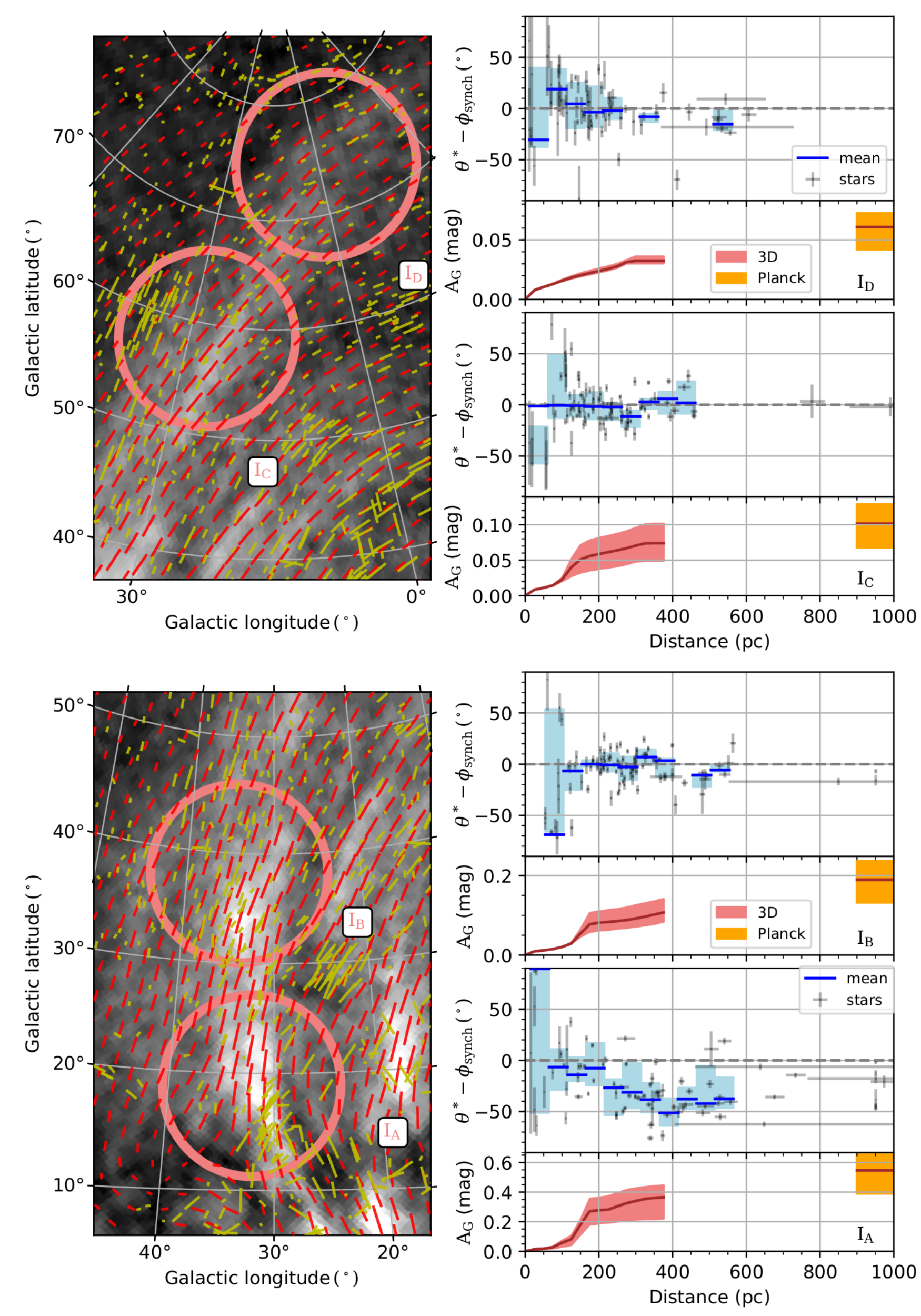}
\caption{Comparison between starlight and synchrotron polarization angles towards Loop\,I. The analysis is performed in four $16^{\circ}$-diameter circular regions ($\rm I_A, I_B, I_C, I_D)$. \textit{Left:} Background: Polarized intensity of synchrotron emission at $N_{\rm side}=128$. Overlay: Stellar polarization segments for stars within 1\,kpc ({\it yellow}) and synchrotron polarization segments ({\it red}) at $N_{\rm side} = 32$. The stellar segment length is scaled by 1/$\sigma_\theta^*$, while the synchrotron segment length is proportional to the polarized intensity. Pink circles mark regions $\rm I_A - I_D$. The map is in Gnomonic projection. The polarization segment visualization is with respect to the direction of North of the center of the image. This does not affect the analysis. \textit{Right:} (top sub-panel) $\theta^* - \phi_{\rm synch}$ as a function of distance (light gray points). Angle differences are binned in distance and we show the mean angle difference in each bin (blue line) and the 16$-$84 percentile range (light blue shade). We do not show means for bins with less than 2 stars. Stars with distances $>1$\,kpc are displayed at 950\,pc for visualization but not included in the analysis. (Bottom sub-panel) Median cumulative $\rm G-$band extinction as a function of distance, evaluated at the coordinates of each star in the circular region, from the 3D dust map of \cite{Leike2020} (red line). The pink range shows the 16- and 84 percentile of $A_{\rm G}$. The total $A_{\rm G}$ derived from {\it Planck} is shown as a solid brown line at large distances (median), with the orange band denoting the 68 per cent range. The angle difference map is at $N_{\rm side}=32$.}
\label{fig:loopI_lower}
\end{figure*}

The synchrotron and thermal dust angles are well aligned for most of the available sky, except for the inner plane ($|l|\lesssim 60^{\circ}$) at intermediate latitudes ($|b|\approx 5^{\circ}$--$20^{\circ}$), and a few other specific regions (e.g., near the Orion nebula). Areas of misalignment can also be informative. In these regions the synchrotron and dust emission are tracing different $B$-fields along the line-of-sight; either due to being at different distances, or a complex geometry producing different projections of the $B$-field on the sky. 

For most of the high-latitude sky (76\% of the unmasked pixels with $|b| > 30^\circ$), the $|\chi_{\rm dust}-\phi_{\rm synch}|$ map shows large, continuous regions with values $\leq 20^{\circ}$. Similar regions of alignment are found in the $|\theta - \chi_{\rm dust}|$ map. This is not a surprise, since at high latitudes, the absorbing and emitting dust column is likely the same \citep[as shown by][]{Skalidis2019,Planck2018_XII}. At high latitudes, the large areas of alignment are particularly above the Galactic centre near Loop\,I. Notable areas of misalignment include  Loops Is and XII where the angles can be discrepant by many tens of degrees. We note that in some of these pixels at high latitudes there is only a single star measurement, which may not be fully representative of the $B$-field along the line-of-sight depending on (a) the distance of the star and (b) whether the source is intrinsically polarized or not. However, in regions where multiple pixels show similar values, this is less of a concern. At low latitudes, the regions with good agreement are likely due to the dominance of the $B$-field aligned with the spiral arms in the Galactic disk.

Our goal is to look for correlations between the various $B$-field tracers and, ultimately, place limits on the distance to the large-scale radio loops/spurs. We therefore expect to be able to perform this correlation analysis for features that meet two conditions:
\begin{enumerate}
    \item[1.] Stellar measurements towards the feature have a high enough density and area coverage.
    \item[2.] Stellar polarization angles are in good agreement with synchrotron polarization angles, and hence can be assumed to be tracing the same $B$-field. 
\end{enumerate}

In practice, we are still very limited by the stellar polarization sampling. Only 2 out of the 14 features studied by \cite{Vidal2015} meet these qualitative criteria. Many loops only have a few stellar measurements that do not cover an appreciable fraction of their extent (IIIs, VIIb, X, XI, XIII, XIV) or have no measurements at all (XII). 

Three loops do have good statistics of stellar measurements but the polarization angles of the stars are misaligned with respect to the synchrotron emission (III, VIII, GCS). This means that the locations of the dominant sources from each tracer, must be different, since they are tracking a different $B$-field direction. This misalignment could in principle be used to gain insight into the statistical variability of the $B$-field and  properties such as field reversals \citep[e.g.,][]{Jaffe2019}. 

Loop Is has tens of measurements within a single \mbox{$N_{\rm side}$ = 32} pixel; these are targeted measurements towards the Corona Australis (CrA) molecular cloud \citep{Heiles2000,Targon2011}, which has no specific relevance to this large-scale loop \citep[see][for a detailed analysis of the magnetic field in this structure]{Bracco2020A&A}. Loop\,IX is in a region of small $\theta^*-\phi_{\rm synch}$, however the coverage of stellar measurements is highly non-uniform, with very few stars coinciding with the ridge of synchrotron polarized intensity. 
 
In the following subsections, we examine in detail the relation between starlight polarization and the synchrotron/thermal dust emission towards the two best-sampled radio spurs: Loop\,I and Loop\,IV. 

\subsection{Loop\,I}
\label{sec:results_loopI}

Loop\,I occupies the region of the high-latitude sky with the best sampling in terms of stellar polarimetry. As seen in Fig.\,\ref{fig:maps}, stellar polarization angles are well-aligned with synchrotron angles in this general area. 
In Fig.\,\ref{fig:loopI_lower} we re-examine the correlation of stellar and synchrotron polarization angles towards Loop\,I in detail. The improvements include more stars, higher S/N radio polarization data, and, most importantly, reliable distance measurements from {\it Gaia}.

We select 4 circular regions ($\mathrm{I_A,I_B,I_C,I_D}$) centered along the outline of Loop\,I that was  defined in \citet{Vidal2015}, to compare the polarization angles of stars and synchrotron emission as a function of stellar distance. The regions are outlined in Figures \ref{fig:loops} and \ref{fig:loopI_lower}. Because of the strong depolarization of the synchrotron emission at low latitude ($b \lesssim 10^\circ$), where line-of-sight confusion is likely prominent, we focus this analysis on areas at higher latitude. For the selected regions we calculate the angle difference of each individual star with the synchrotron polarization angle of the $N_{\rm side}=32$ pixel it occupies. We choose a radius of 8$^\circ$, wide enough to contain a large stellar sample, but small enough to avoid overlap with other features of the radio sky. The four regions are centered on the loop at increasing latitudes; their coordinates can be found in Table\,\ref{tab:regions}. In each region, we bin the stellar data in distance using a constant distance range of 50\,pc. The first bin edge is placed at the distance of the nearest star. In each bin we calculate the weighted mean angle difference using equation \ref{eqn:circmean} and quantify the spread as the 16-84 percentile range. These values are shown along with the individual stellar measurements in Fig.\,\ref{fig:loopI_lower} (upper sub-panels for each region). Bins with less than 2 stellar measurements are excluded from this calculation. Stars further than 1\,kpc are not used, as they would not provide further constraining capability (the angle differences have already converged at much smaller distances, as seen in Fig. \ref{fig:loopI_lower}).

\begin{table}
\centering
\caption{Summary of location and distance measurements for the Loop\,I and IV regions.}
\begin{tabular}{lcc}
\hline
Region		                &Location                   &Distance limits         	\\
                            &$(l,b)$                    &(pc)               \\
\hline
$\mathrm{I}_{\mathrm{A}}$   &$(32^{\circ}\!.10, 19^{\circ}\!.12)$ & --       \\
$\mathrm{I}_{\mathrm{B}}$   &$(33^{\circ}\!.95, 38^{\circ}\!.03)$ & $112 \pm 17$\,pc   \\
$\mathrm{I}_{\mathrm{C}}$   &$(26^{\circ}\!.68, 58^{\circ}\!.73)$ &$\leq 105$\,pc        \\
$\mathrm{I}_{\mathrm{D}}$   &$(358^{\circ}\!.17, 74^{\circ}\!.06)$ &$135 \pm 20$\,pc  \\
\hline
$\mathrm{IV}_{\mathrm{A}}$   & $(328^{\circ}\!.74, 33^{\circ}\!.59)$ &$180 \pm 15$\,pc   \\
$\mathrm{IV}_{\mathrm{B}}$   &$(302^{\circ}\!.72, 31^{\circ}\!.81)$ & --         \\
\hline
\end{tabular}
\label{tab:regions}
\end{table}
\normalsize

In all regions we find that the nearest distance bins spanning [0, 100]\,pc, contain measurements with large uncertainties in $\theta^*-\phi_{\rm synch}$. The uncertainties of the angle differences at these distances are dominated by the uncertainties of the stellar measurements, $\sigma_\theta^*$. The mean $\sigma_\theta^*$ is $\sim 20^\circ$ for regions $\rm I_B,I_C,I_D$ in this range and 12$^\circ$ for region $\rm I_A$. The spread of the distribution of $\theta^*$ is also large. At larger distances, the stellar uncertainties decrease significantly (by about a factor of 3--5).

This behavior arises from the distribution of ISM dust within the vicinity of the Sun. We are located in a cavity of the ISM, known as the Local Bubble, where the dust volume density is low \citep{Cox1987}. This cavity has recently been mapped with the use of stellar extinctions and distances \citep{Lallement2018,pelgrims2020}. To confirm that indeed the observed large uncertainties at small distances are due to a relative absence of dust, we employ the 3D dust extinction map of \cite{Leike2020}.

We query the map to obtain the extinction profile at the coordinates of every star in our selected regions. Figure \ref{fig:loopI_lower} (lower right sub-panels) shows the median, 16 and 84 percentiles of the distribution of $G$-band extinction towards the stars, $A_{\rm G}$, as a function of distance. An abrupt rise in $A_{\rm G}$ is seen at distances $\sim$ 100\,pc, in good agreement with the range of distances where we observe large $\sigma_\theta^*$. The rise in extinction coincides with a rise in the stellar polarization at these distances, as shown in \citet{Berdyugin2014} for the high-latitude parts of Loop\,I and in \citet{Santos2011} for the lower latitude area (our region $\mathrm{I_A}$).

The mean values of $\theta^*-\phi_{\rm synch}$ in the distance range [100, 200]\,pc, are within $\pm10^\circ$ for regions $\rm I_B, I_C, I_D$.  Within this distance range, the two tracers are probing the same orientation of the magnetic field. This remains the case at larger distances (out to 600\,pc). 

The lowest latitude region, $\rm I_A$, shows a different behavior. While the mean angle differences are within 20$^\circ$ of the origin for the distance range [100, 200]\,pc, the standard deviation is large due to the substantial intrinsic scatter of the measurements. The mean angle between $\theta^*$ and $\phi_{\rm synch}$ deviates gradually from zero at distances greater than $\approx 250$\,pc, dropping from $-5^{\circ}$ to $\approx -40^\circ$ at distances larger than 300\,pc. The spread of the distribution of angle differences is large, and is dominated by astrophysical scatter; the uncertainties of individual $\theta^*$ are on average $< 5^\circ$. At these distances, stars in region $\mathrm{I_A}$ are probing a magnetic field geometry that is different than that probed by the synchrotron emission. This is not unexpected since this region is the lowest latitude portion of Loop\,I studied here and therefore the line-of-sight magnetic field will be more complex. The dust and magnetic field responsible for polarizing stars further than 200\,pc may be associated with the outskirts of the Aquila Rift, which extends over tens of square degrees to the right of the map and lies at a distance of 200--280\,pc \citep{zucker2020}.

So far, we have found an alignment between the mean stellar polarization angle and the magnetic field orientation traced by the synchrotron emission in the distance range [100, 200]\,pc for $b > 30^\circ$. This is a strong indication that at high latitude, Loop\,I, which dominates the polarized synchrotron emission in the area, is located within this range of distances. This conclusion is in agreement with several other determinations of the distance to Loop\,I \citep[e.g.,][]{Salter1983,Vidal2015,Planck2015_XXV,Das2020}. 

\begin{figure}[ht]
\centering
\includegraphics[scale=1]{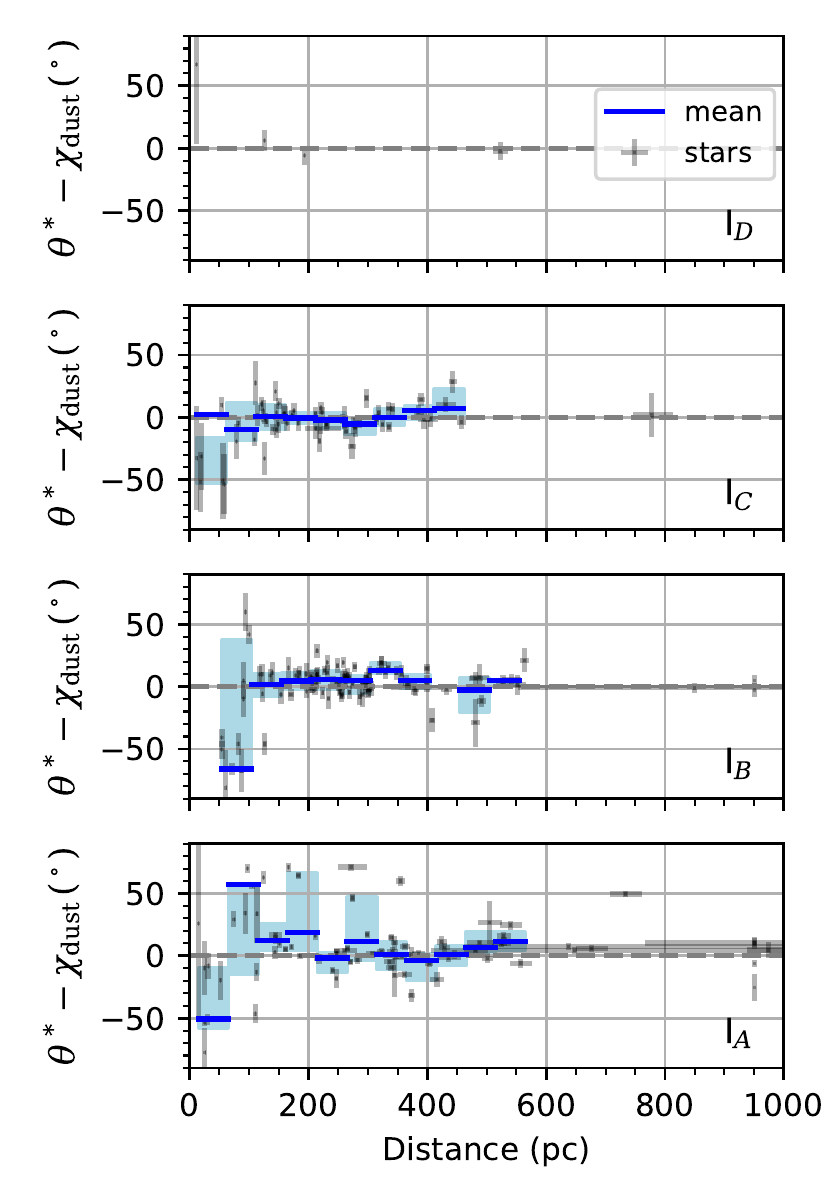}
\caption{Difference between stellar and thermal dust $B$-field angles, $\theta^*$ and $\chi_{\rm dust}$, in degrees, as a function of distance. The four panels represent the four regions selected for Loop\,I. Symbols are the same as in Fig.\,\ref{fig:loopI_lower}.}
\label{fig:dust_stars_I}
\end{figure}

\begin{figure*}[ht]
\centering
\includegraphics[scale=0.95]{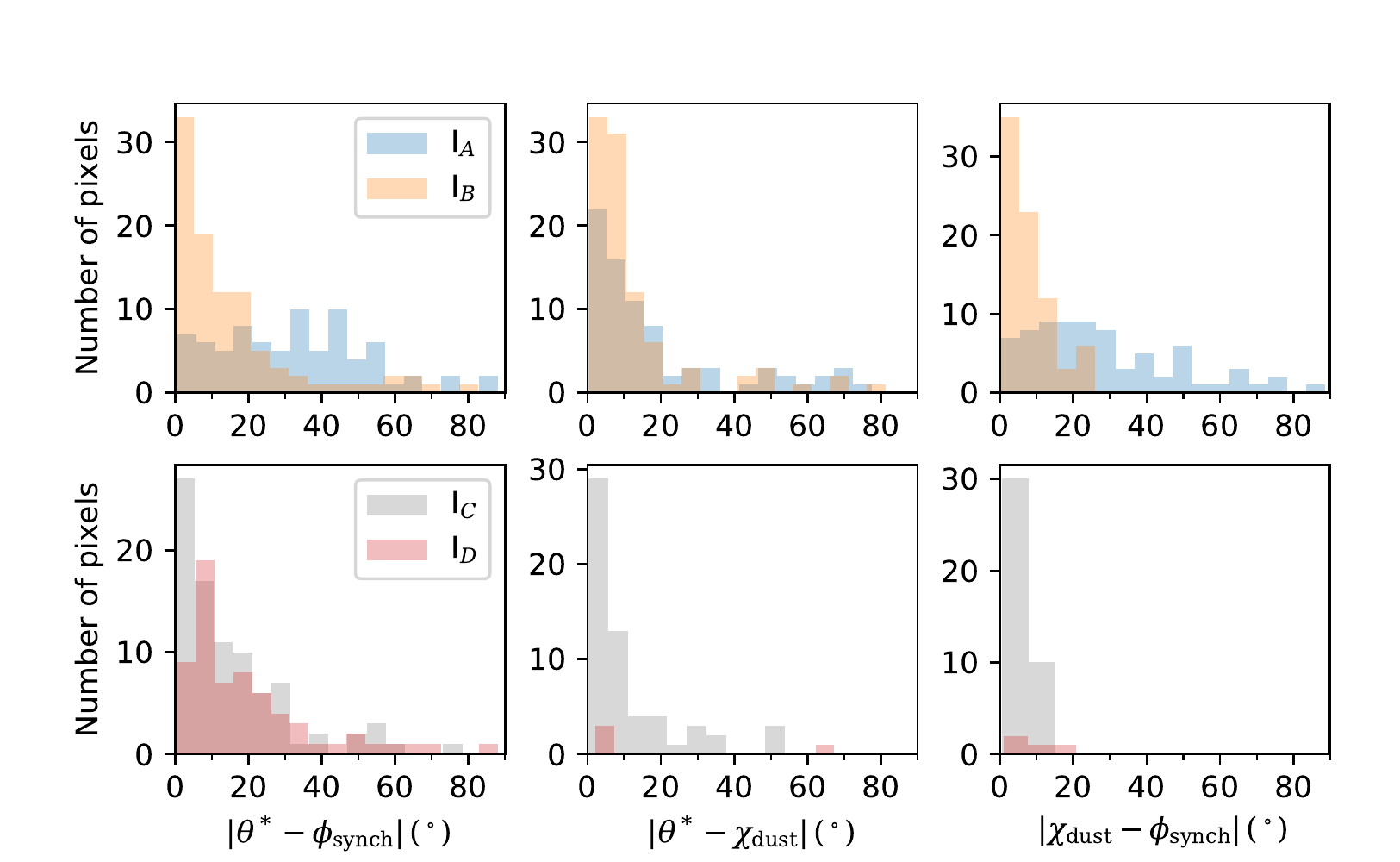}
\caption{Distributions of polarization angle differences in degrees between the maps of the three magnetic field tracers for the 4 regions centered on Loop I (see Fig.\,\ref{fig:loopI_lower}). All non-masked pixels within the 4 circular regions are included in the calculation of the distribution. The blue, orange, gray and red histograms correspond to regions $\mathrm{I_A}$, $\mathrm{I_B}$, $\mathrm{I_C}$ and $\mathrm{I_D}$, respectively.}
\label{fig:loopI_distributions}
\end{figure*}

In our analysis, we have calculated the angle difference between the synchrotron emission measured within $1.\!^{\circ}8$ pixels, and individual stellar polarizations (measured within the arcsecond stellar PSF). The low spatial density of stellar measurements has necessitated this choice. However, it is natural to expect that this mismatch of spatial scales may contribute to the observed astrophysical scatter in the angle differences. 
If this is the case, we expect that the distribution of angle differences, if computed at the same resolution for both tracers, would show a tighter correlation than the distribution using individual stellar measurements. Though limited by the amount of stellar polarization data, we can gain further insight through the tight correlation between dust-induced polarization in absorption and in emission that holds at high Galactic latitude \citep{Planck2018_XII}.

First, we search for the distance at which the correlation between dust emission and stellar polarization angles occurs \citep[as in][]{Skalidis2019}. Figure~\ref{fig:dust_stars_I} shows the angle difference $\theta^*-\chi_{\rm dust}$ as a function of distance for the four regions along Loop\,I. The two tracers are well-aligned for distances larger than 100\,pc in regions $\mathrm{I_B}$, $\mathrm{I_C}$, and $\mathrm{I_D}$. The statistics in region $\mathrm{I_D}$ are poor due to the low S/N of the {\it Planck} polarized dust emission data at high latitude, however all $\theta^*$ in the region lie within $1\,\sigma$ of $\chi_{\rm dust}$. From the observed tight correlation we infer that the polarized dust emission is dominated by the dust at 100\,pc in these regions. 

While the polarized dust emission is mostly local in these high latitude regions, the cumulative extinction (Fig.\,\ref{fig:loopI_lower}) appears to rise out to the maximum distance of the 3D dust map ($\sim 400 $\,pc). We can estimate the maximum amount of stellar polarization that this rise in extinction could produce, from the relation between reddening, $E(B-V)$, and maximum stellar polarization fraction, $p_{\rm max}$: $ p_{\rm max} = 13\% \, E(B-V)$ \citep{Panopoulou2019, Planck2018_XII}. For regions $\rm I_C$ and $\rm I_D$, the rise from 150--400\,pc would correspond to a change in fractional polarization of the stars of $<$0.06\%, significantly less than the mean polarization of the stars in the region (0.15\% and 0.3\% for region $\rm I_D$ and $\rm I_C$, respectively). In region $\rm I_B$ the rise from 150--400\,pc is $\sim 0.1$\,mag, which would correspond to a polarization of $0.3\%$, comparable to the mean polarization of stars in the region. The fact that we do not detect significant changes in the polarization angle suggests that the magnetic field geometry does not vary along the line-of-sight, even though the amount of dust (and polarization) may be increasing with distance. 

We do note, however, that the observed rise in 3D extinction may be affected by systematic uncertainties. The median extinction from the 3D dust map does not reach the \textit{Planck} value, which is probing the entire dust column. This discrepancy may be pointing to systematic offsets between the two datasets, which are constructed from entirely different tracers and methods. In fact, at high latitude, existing dust maps show significant systematic variations even among maps created using \textit{Planck} data alone \citep[see, e.g.,][]{Panopoulou2019}. A detailed comparison between the 3D dust map of \citet{Leike2020} and the FIR-inferred extinction from \textit{Planck}, though warranted, is beyond the scope of this paper.

The results in regions $\rm I_B,I_C,I_D$ are very different from those in region $\rm I_A$. In this region the mean angle difference drops below 20$^\circ$ from 0$^\circ$ at $\sim 200$\,pc, however a large spread remains (Fig.\,\ref{fig:dust_stars_I}). This spread, along with the rise in $A_{\rm G}$ in the region farther than 200\,pc may indicate the presence of multiple dust structures with varying magnetic field geometries. At distances larger than 300\,pc, the majority of stars have a $\theta^*$ within 16$^\circ$ of $\chi_{\rm dust}$ and the median $A_{\rm G}$ profile flattens. These indications of the presence of material extended along the line-of-sight agree with similar conclusions from X-ray absorption studies at even lower latitude \citep{Sofue2015,Lallement2016}.

Figure\,\ref{fig:loopI_distributions} shows the distribution of angle differences between all three pairs of $B$-field tracers towards Loop\,I: $|\theta^*-\phi_{\rm synch}|$,  $|\chi_{\rm dust}-\phi_{\rm synch}|$, $|\theta^*-\chi_{\rm dust}|$. The distribution of angle differences takes into account all non-masked pixels within the circular regions defined in Fig.\,\ref{fig:loopI_lower}. As seen previously, stellar polarization in the higher latitude regions $\mathrm{I_B}$, $\mathrm{I_C}$ and $\mathrm{I_D}$ is well aligned with $\phi_{\rm synch}$. This is not the case in the lowest latitude region $\mathrm{I_A}$, where we have established the contribution to the polarization of dust structures further than 200\,pc (Fig.\,\ref{fig:dust_stars_I}). In all regions, stellar polarizations are well-aligned with $\chi_{\rm dust}$, as shown over most of the sky by {\it Planck} \citep{PIP_XXI,Planck2018_XII}, confirming that we can use dust emission to examine the effect that beam-averaged stellar polarizations would have. In regions $\mathrm{I_B}$ and $\mathrm{I_C}$, we find the standard deviation of the distribution of $\theta-\phi_{\rm synch}$ (16$^\circ \pm1^\circ$ and 24$^\circ \pm 3^\circ$) to be larger than that of the $\chi_{\rm dust} - \phi_{\rm synch}$ distribution (9$^\circ \pm 0.^\circ 5$ and 9$^\circ \pm 1^\circ$), as expected due to beam averaging. The standard deviation of the distribution of angle differences and its uncertainty were calculated from 1,000 Monte Carlo realizations, from which we report the mean and $1-\sigma$ uncertainty. Region $\mathrm{I_D}$ (highest latitude region) has very few pixels with high enough S/N in the dust emission map. 

The observed tight correlation between $\chi_{\rm dust}$ and $\phi_{\rm synch}$ in regions $\mathrm{I_B}$ and $\mathrm{I_C}$ lends additional support to the argument that the bright edge of Loop\,I lies at a distance of 100--200\,pc. The dominant contribution to the synchrotron emission along the line-of-sight most likely arises from within the same volume that dominates the polarized dust emission (Fig.\,\ref{fig:dust_stars_I}).

In conclusion, we present two pieces of evidence in support of the hypothesis that at $b > 30^\circ$, Loop\,I is a nearby structure (within $100$--200\,pc): 
\begin{itemize}
    \item The mean stellar polarization, averaged within regions $\mathrm{I_B}$, $\mathrm{I_C}$, $\mathrm{I_D}$ is aligned with the synchrotron magnetic field orientation within 100--200\,pc. 
    \item The dust emission polarization is well aligned with synchrotron in the regions where the total dust column is dominated by structures within a few hundred\,pc.
\end{itemize}

In the analysis so far we have presented a qualitative discussion of the behavior of the values of $\theta^*-\phi_{\rm synch}$. We quantify the uncertainty in the distance determination in Section \ref{sec:limits}. We further discuss how the distance determination in this paper compares to others in the literature in Section \ref{sec:discussion}.

\subsection{Loop IV}
\label{sec:results_loopIV}

\begin{figure*}[ht]
\centering
\includegraphics[scale=0.9]{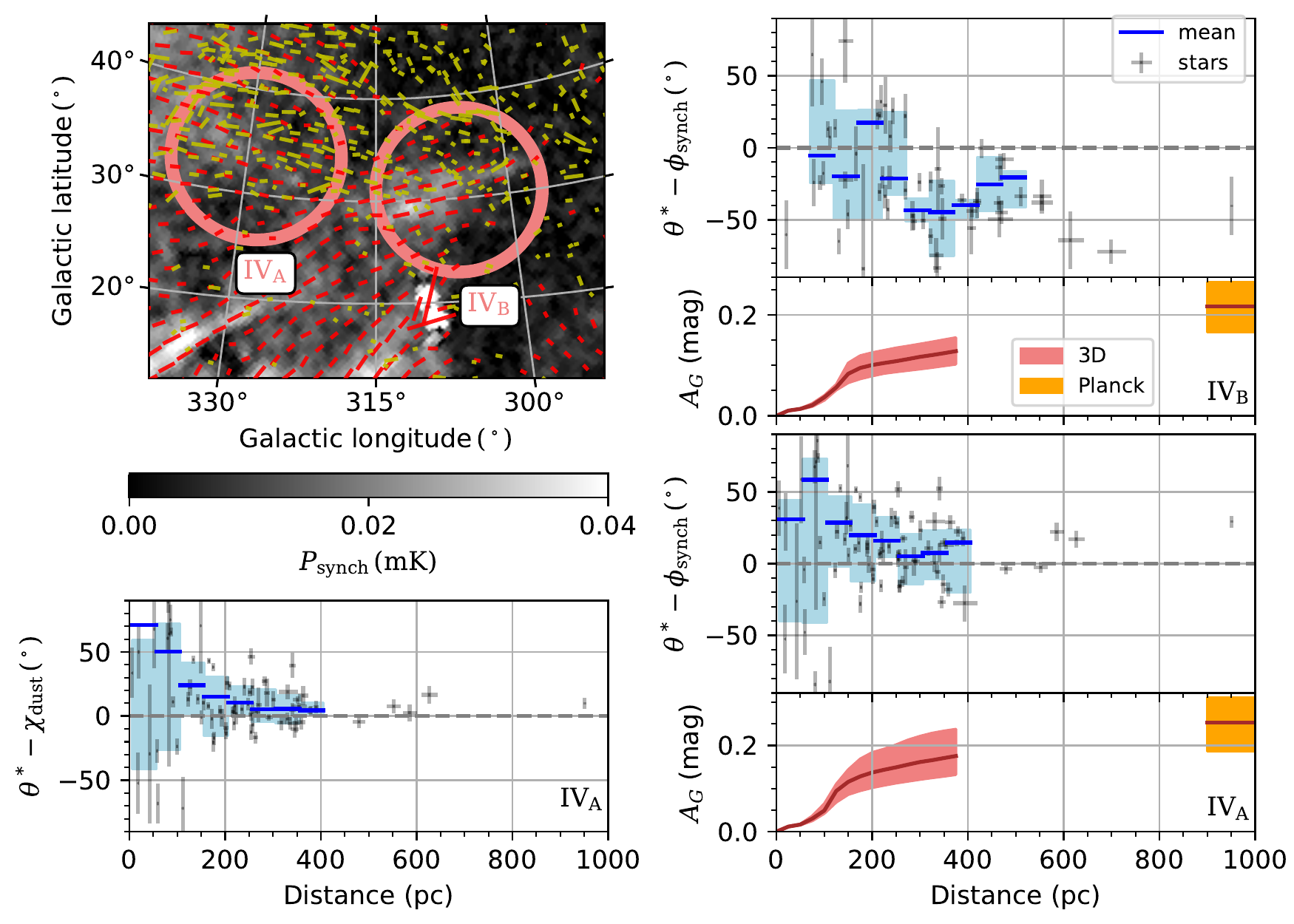}
\caption{As in Fig.\,\ref{fig:loopI_lower} but for Loop\,IV and for two regions (IV$_\textrm{A}$ and IV$_\textrm{B}$). The lower left panel shows the difference between stellar polarization, $\theta^*$, and dust emission, $\chi_{\rm dust}$, as a function of distance, as in Fig. \ref{fig:dust_stars_I} but for region $\mathrm{IV_A}$.}
\label{fig:LoopIV}
\end{figure*}

We next investigate the stellar polarization towards Loop\,IV (Fig.\,\ref{fig:LoopIV}). We perform the analysis of angle differences in two circular regions of radius $8^\circ$, centered on the outline of Loop\,IV that was  defined in \citet{Vidal2015}. In both regions the binned angle difference $\theta^* - \phi_{\rm synch}$ is significantly offset from zero for the majority of distance bins. The lowest (absolute) values of mean angle difference are found in region $\mathrm{IV_A}$ within 250--350\,pc. We are faced with two apparently conflicting results: stellar and synchrotron polarizations are indicative of alignment in this distance range for region $\mathrm{IV_A}$, but no such alignment is observed for its neighboring region $\mathrm{IV_B}$. These results suggest two possibilities: (a) that Loop\,IV is located further than $\sim 400$--500\,pc, where stars are lacking from our analysis, in which case the alignment in region $\mathrm{IV_A}$ at 250--350\,pc is circumstantial, or, (b) synchrotron emission from regions $\mathrm{IV_A}$ and $\mathrm{IV_B}$ originates at drastically different distances.

We can test the first hypothesis by looking at the dust emission towards this structure. First, we examine the dependence of $\theta^*-\chi_{\rm dust}$ in region $\mathrm{IV_A}$ (bottom left panel, Fig.\,\ref{fig:LoopIV}). The angle differences $\theta^*-\chi_{\rm dust}$ converge to zero for distances larger than 300\,pc. The polarized dust emission is therefore dominated by dust within 300\,pc in region $\mathrm{IV_A}$. In region $\mathrm{IV_B}$ there are too few pixels with both stellar measurements and high S/N polarized dust emission at $N_{\rm side}= 32$. We confirm that at $N_{\rm side}=16$, the mean difference $\theta^*-\chi_{\rm dust}$ approaches zero at a distance of 300\,pc, though the independent pixels with $\chi_{\rm dust}$ measurements are only 10 (not shown in Fig.\,\ref{fig:LoopIV}).

\begin{figure*}[ht]
\centering
\includegraphics[scale=1]{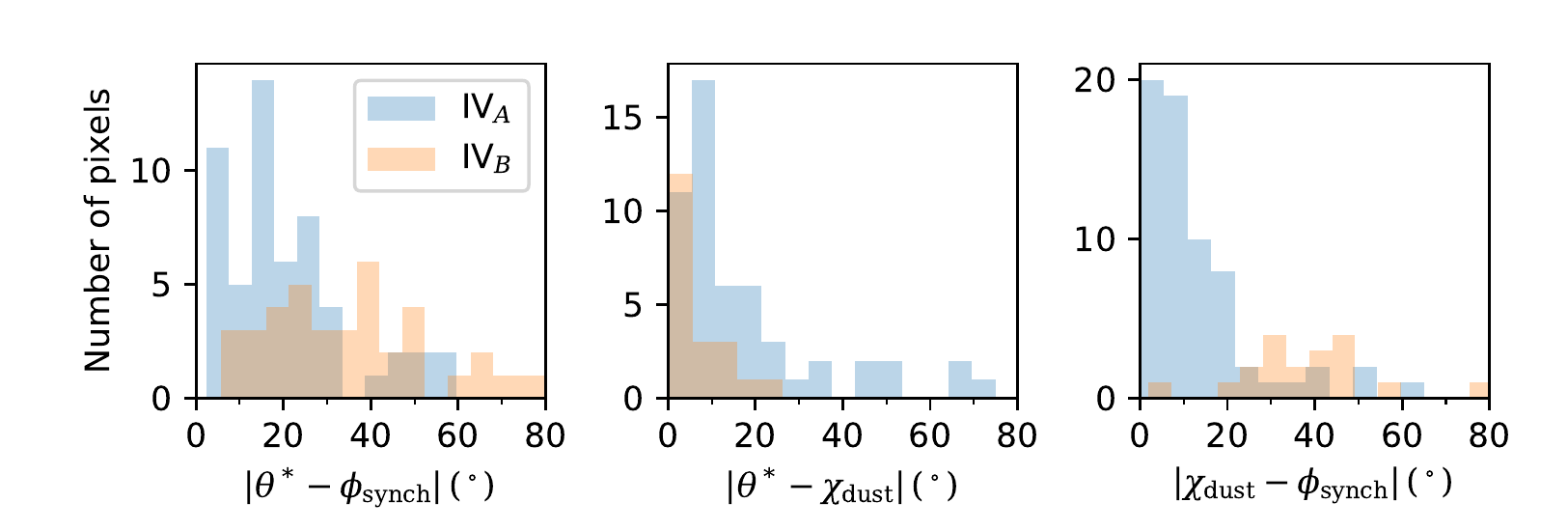}
\caption{Distributions of polarization angle differences between the three magnetic field tracers for each of the 2 regions centered on Loop IV (see Fig. \ref{fig:LoopIV}).}
\label{fig:loopIV_distr}
\end{figure*}

\begin{figure*}[ht]
\centering
\includegraphics[scale=0.93]{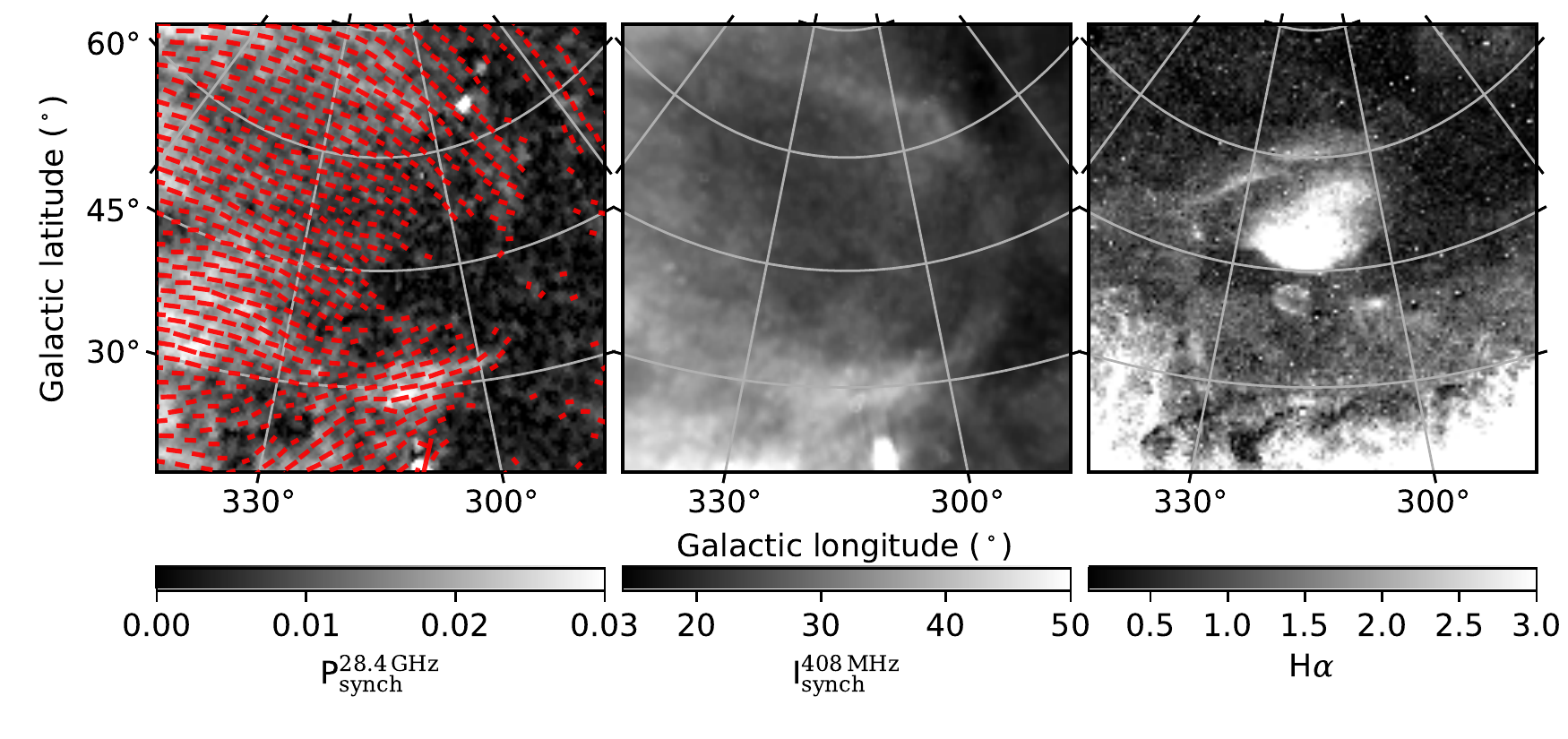}
\caption{Maps showing the entire Loop\,IV, in Gnomonic projection. \textit{Left}: Synchrotron polarized intensity at 28.4\,GHz with polarization segments parallel to the magnetic field overlaid. \textit{Middle}: Synchrotron total intensity at 408\,MHz where the loop outline is most clear. \textit{Right}: Map of $H\alpha$ emission showing the Spica \ion{H}{2} region, most likely unrelated to Loop\,IV (see text). }
\label{fig:loopIV_tracers}
\end{figure*}

Next, we examine the distribution of absolute angle differences between pairs of all three magnetic field tracers (Fig.\,\ref{fig:loopIV_distr}). 
For region $\mathrm{IV_A}$, we find that the dust emission is tightly aligned with $\phi_{\rm synch}$, more so than the stellar measurements. The standard deviation of the distribution of $\theta^*-\phi_{\rm synch}$ is 25$^\circ \pm 2^\circ$, versus 19$^\circ \pm 1^\circ$ for the distribution of $\chi_{\rm dust}-\phi_{\rm synch}$ (where we report the mean and $1\sigma$ from Monte Carlo realizations taking into account the measurement uncertainties). We have established from our discussion of Fig.\,\ref{fig:LoopIV} that the polarized dust emission predominantly arises from within 300\,pc. From their tight correlation, we conclude that the synchrotron polarization in region $\mathrm{IV_A}$ must be within 300\,pc, co-located with the dust emission. However, the same situation does not hold for region $\mathrm{IV_B}$. In this region, despite the fewer detections of $\chi_{\rm dust}$, it is clear that the $\chi_{\rm dust}-\phi_{\rm synch}$ show a large spread, with no preferred alignment at $0^{\circ}$. A comparison of stellar polarizations with dust emission, suggests that in this region the polarized dust emission arises from within 300\,pc (as in region $\mathrm{IV_A}$). Since the dust emission is not aligned with $\phi_{\rm synch}$ here, the synchrotron emission in region $\mathrm{IV_B}$ cannot be at the same distance as that in $\mathrm{IV_A}$. It would have to be located further than the stars in our sample, i.e., further than 550\,pc. Otherwise, it would need to have no dust associated with it. In both cases, we can deduce that $\mathrm{IV_A}$ and $\mathrm{IV_B}$ are not part of the same physical structure -- the data favor hypothesis (b).

To the best of our knowledge, this is the first conclusion in the literature that Loop\,IV is not a coherent structure. The fact that $\phi_{\rm synch}$ are misaligned with the axis of the structure in region $\rm IV_A$ \citep[as shown by][]{Vidal2015} supports this conclusion. We note that Loop\,IV was first identified in total intensity \citep{Large1966,Berkhuijsen1971} and was initially thought to be part of Loop\,I. Figure\,\ref{fig:loopIV_tracers} shows maps of synchrotron total intensity at 408\,MHz from \citet{Remazeilles2015} as well as the map of polarized intensity at 28.4\,GHz from this work. We first note that the apparent loop in intensity, which is where it is most clear, is only visible on one side. Furthermore, the upper part is actually co-located with the `tip' of Loop\,I where it fades into the background. Therefore, a large part of the circular feature may be due to chance alignment, especially given the morphology in polarization.

In polarized intensity the outline of Loop\,IV is not as continuous as in total intensity, with large gaps along it (see Fig.\,\ref{fig:loopIV_tracers}). The polarization angles do not trace the outline of the loop over most of its length \citep[see also][]{Vidal2015}. Rather, the polarization angles appear as a continuation of the large-scale pattern that exists throughout a much larger area outside the loop (towards the left of the map). When comparing the polarization angle with the loop's outline in the lower part of Loop\,IV, \citet{Vidal2015} find that the two are aligned only over the part of the structure that we define as region IV$\rm _B$.

We speculate that the polarized intensity in region $\mathrm{IV_B}$ could be associated with a different structure than what has been traditionally defined as Loop IV. Upon inspection of the map in Fig.\,\ref{fig:LoopIV}, we observe that the polarization pattern of the synchrotron emission in region $\mathrm{IV_B}$ appears as a smooth continuation of the upper part of Loop\,XIV (bright filament in the lower left corner of the map). Targeted observations of stellar polarization towards the region connecting these two structures would help test this hypothesis, and yield a distance limit to the associated structures.

As a final note, we mention the existence of the bright bubble  in $\rm H\alpha$ discovered by \citet{Reynolds1985} (shown in the bottom panel of Fig.\,\ref{fig:loopIV_tracers}), which appears co-located with the center of Loop\,IV. This \ion{H}{2} region is associated with the bright binary star Spica ($\alpha$\,Vir). The \ion{H}{2} region is also observed in FUV spectral line emission \citep{Park2010,Choi2013}. With a \textit{Hipparcos} parallax of $12.44 \pm 0.86$ mas \citep{Hipparcos1997}, the distance of Spica is $80\pm 5$\,pc.  The \ion{H}{2} region is therefore much nearer than where the synchrotron emission originates from in region IV$\rm_A$, hence the two structures are most likely unrelated.

\subsection{Calculating upper and lower distance limits for Loops I and IV}
\label{sec:limits}

Having discussed the qualitative behavior of $\theta^*$ and $\phi_{\rm synch}$ as a function of distance for the selected regions towards Loops I and IV (Sections \ref{sec:results_loopI} and \ref{sec:results_loopIV}), we proceed to place lower and upper limits on the distance to the synchrotron emission. 

If $\theta^*$ and $\phi_{\rm synch}$ are tracing the same magnetic field at a distance $D_{\rm synch}$, we expect to observe mean angle differences $\theta^*-\phi_{\rm synch}$ around zero for distances further than $D_{\rm synch}$. This places an \textit{upper limit} on the distance of the synchrotron emission. In practice, we search for distance bins where two criteria for alignment are met:
\begin{itemize}
    \item[1.] The absolute value of the circular mean of $\theta^*-\phi_{\rm synch}$ in a bin is $\leq 20^\circ$ and
    \item[2.] The PRS within the bin is higher than the 99.9 percentile of the distribution of PRS drawn from a uniform distribution (equivalent to a significance threshold of 3$\sigma$).
\end{itemize}
As the various sources of noise (measurement uncertainty, intrinsic astrophysical scatter) can cause spurious cases in which these criteria are met, we search for distance ranges where the conditions are met over multiple bins.

\begin{figure*}[ht]
\centering
\includegraphics[scale=1]{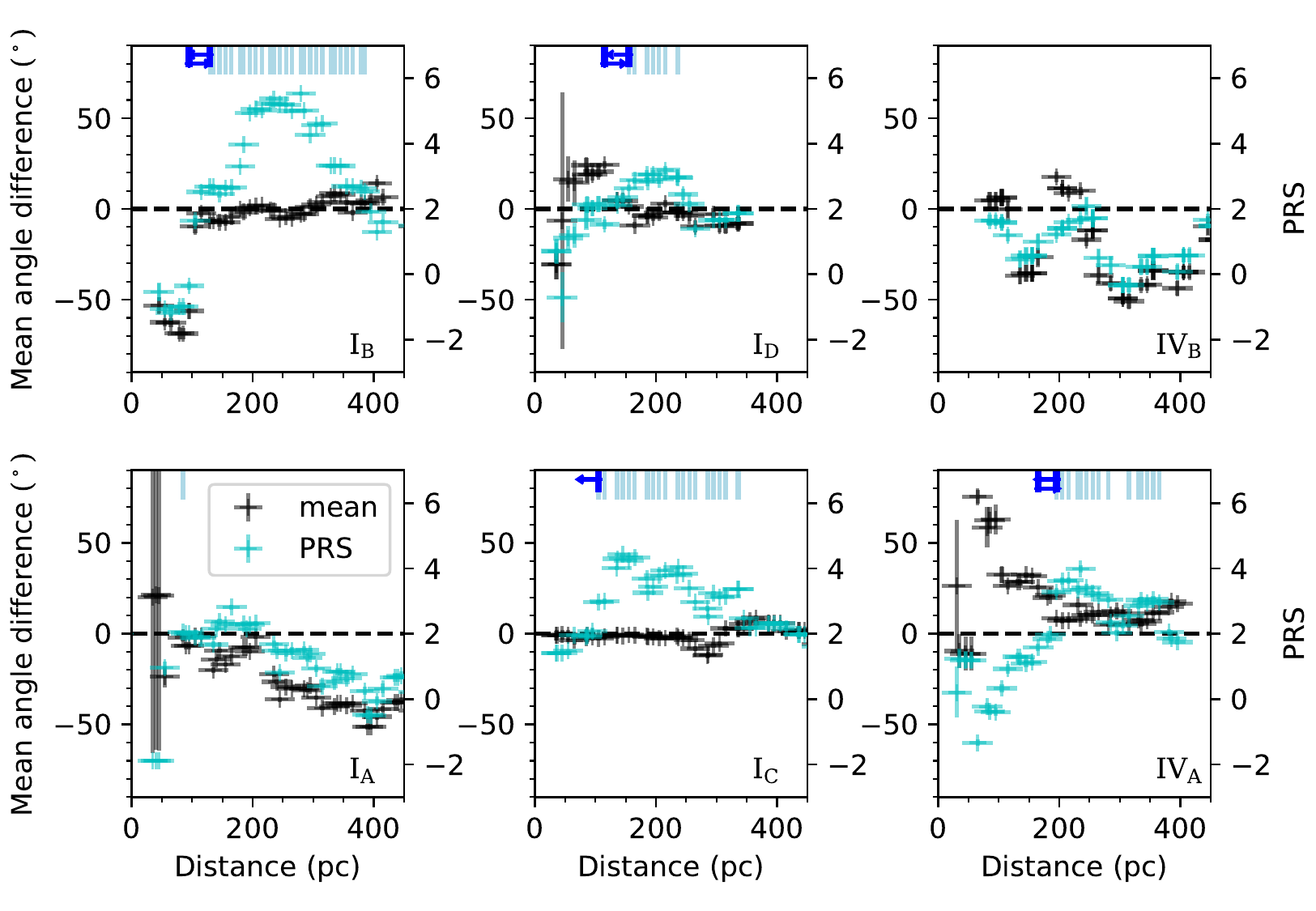}
\caption{Distance-binned statistics of $\theta^* - \phi_{\rm synch}$ used for the determination of the distance to synchrotron features in regions $\rm I_A, I_B, I_C, I_D, IV_A, IV_B$ (see labels). Symbols show the circular mean (black crosses) and PRS (cyan crosses) of the distribution of angle differences $\theta^* - \phi_{\rm synch}$ within distance bins of 50\,pc width. Error bars denote the $\pm 1$ standard deviation from 10,000 Monte Carlo realizations. The light blue lines at the top of the panels mark the centers of bins where the conditions for alignment are fulfilled. Upper and lower limits for the distance to the synchrotron features are shown with blue arrows.}
\label{fig:appendix}
\end{figure*}

Placing a lower limit on the distance is less straightforward, especially if there is significant contribution from $B$-field foreground to $D_{\rm synch}$. Fortunately, we have established that for regions $\rm I_B, I_C, I_D$ there is little foreground contamination up to the distance where we observe alignment (dust extinction is low at distances $<100$\,pc). Region $\rm IV_A$ also exhibits alignment after some distance, even though the dust extinction continues to rise up to 400\,pc, suggesting that the foreground dust is not dominating the polarization signal. 
To place a \textit{lower limit} on the distance to the synchrotron features, we can use the fact that the mean $\theta^*-\phi_{\rm synch}$ will in general be non-zero at distances where the stars are tracing a different magnetic field than the synchrotron. We search for bins where the mean $\theta^*-\phi_{\rm synch}$ shifts from significantly non-zero values to being consistent with zero. Specifically, distance bin $i$ will fit the criteria for a lower distance limit if:

\begin{equation}
    \mu^{i+1} \leq 20^\circ \,\,\,{\rm and}\,\,\, \mu^{i+1} - \mu^i \geq 2\sigma_{\mu^i},
\end{equation}
where $\mu^i$ is the mean value of the $\theta^*-\phi_{\rm synch}$ in distance bin $i$, and $\sigma_{\mu^i}$ is the uncertainty on $\mu^i$, calculated as the standard deviation from 10,000 random realizations of the data in the bin (the same realizations used to obtain uncertainties in the PRS, discussed below). Obviously, the center of bin $i$ must be at a smaller distance than the upper limit on the distance, as defined by the criteria (1) and (2) mentioned previously.

Before we proceed, we must take into account the following sources of uncertainty that were ignored in the analysis of Sections \ref{sec:results_loopI} and \ref{sec:results_loopIV}: (a) the selection of distance bins, and (b) the statistical significance of the alignment, as quantified by the PRS.

\textit{Placement of distance bins.}
To quantify whether stellar and synchrotron polarization angles are statistically aligned at a certain distance, we bin the $\theta^*-\phi_{\rm synch}$ measurements in distance steps of 50\,pc. In Sections \ref{sec:results_loopI} and \ref{sec:results_loopIV} we showed one such choice of bins. We investigate the effect of shifting the bin locations on the observed correlations as follows. Starting from a distance of 0\,pc, we define bins of size 50\,pc out to a distance of 1\,kpc. Thus, the first bin spans [0, 50]\,pc, the second [50, 100]\,pc etc. We compute the circular mean and PRS in each bin that contains more than 2 stars. We repeat this process by shifting the bins by 10, 20, 30 and 40\,pc. The measured circular mean and PRS in all bins are shown in Fig.\,\ref{fig:appendix}. 

\textit{Measurement uncertainties and the significance of the PRS.}
We explore the effect of uncertainties of individual measurements as follows. We `re-oberve' each angle difference measurement by drawing from a normal distribution with mean equal to the observed angle difference $\theta^*-\phi_{\rm synch}$ and standard deviation equal to the measurement uncertainty. We repeat this process 10,000 times for each bin. We recompute the mean angle difference and PRS for each iteration. The standard deviation of the values is the error on the mean angle difference and PRS in each bin. 

To quantify the statistical significance of the alignment, we compute the PRS for random realizations of measurements drawn from a uniform distribution \citep[][]{Jow2018}. Each mock angle difference is drawn from a normal distribution with mean a random value in the range [-90$^\circ$, 90$^\circ$] and standard deviation equal to the measurement uncertainty. If the observed PRS value is larger than the 99.9 percentile of mock (uniform) PRS values in that bin, then the alignment is statistically significant (at a level of $\geq 3\,\sigma$).

\textit{Distance determination.}
The results in Fig.\,\ref{fig:appendix} confirm the qualitative result that alignment of $\theta^* - \phi_{\rm synch}$ is found for regions $\rm I_B, I_C, I_D, IV_A$ and not for regions $\rm I_A, IV_B$. Region $\rm I_A$ shows an example of spurious alignment in Fig.\,\ref{fig:appendix} for the bin centered on 85\,pc. Neighboring bins do not meet both alignment criteria.

Region $\rm I_B$ shows alignment for distance bins over a wide range: the alignment conditions are met for the bins centered at 130\,pc out to 385\,pc. Thus, the upper limit to the distance of the synchrotron emission in the region is 130\,pc.
The lower distance limit criterion is met for the bin centered on 95\,pc. We thus conclude that the synchrotron emission from region $\rm I_B$ must arise within the range $[95, 130]$\,pc. Since the bin spacing is 10\,pc, these values may vary within $\pm 10$\,pc. 

We follow the same process to place distance limits for region $\rm I_C$. The distance bin centered on 105\,pc is the first to meet the alignment criteria. The alignment criteria are met for bins spanning [105, 335]\,pc. At small distances, the PRS is within 3$\sigma$ of the uniform case, indicating that the observed small values of the mean angle difference are not statistically significant. The fact that we do not observe a shift in the mean values with distance precludes the determination of a lower distance limit. We do note that the low PRS values coincide with a distance range devoid of dust (see Section\,\ref{sec:results_loopI}). This distribution of dust with distance would suggest that the dust causing the polarization of stars is beyond 55\,pc. The synchrotron emission in region $\rm I_C$ likely exists within the range [55, 105]\,pc. However, the stellar polarization data alone only allow us to place confidence in the upper limit of 105\,pc.

In region $\rm I_D$ the alignment criteria are met in the narrow range of distances for bins centered at [155, 235]\,pc. The values of the PRS are lower compared to other regions, due to the increased uncertainty in the stellar polarization measurements. We can place a conservative upper limit on the distance to the synchrotron emission in region $\rm I_D$ of $ \leq 155$\,pc. The lower limit is found at 115\,pc. Thus, the distance to the synchrotron emission in region $\rm I_D$ is [115, 155]\,pc. A larger number of more precise measurements would be needed to give tighter constraints.

We have attempted to combine the stellar measurements in all three regions of Loop\,I with distance constraints, in order to increase the statistics of the number of stars and obtain a more precise measurement of the distance. However, combining these measurements results in much increased scatter per distance bin. 
This increased scatter may be related to the fact that there seems to be a distance gradient with latitude, as shown by the stellar extinction measurements of \citet{Das2020}. These authors find that the distance to the dust layer associated with Loop\,I varies gradually with latitude from 131 $\pm 7$\,pc (at $b = 32^\circ$) to 70 $\pm$ 4\,pc (at $b=55^\circ$). 

Region $\rm IV_B$ shows no bin with alignment. The PRS is low throughout the observed distance range, suggesting that along the sightline there is no dust feature that is co-located with the synchrotron emission and would cause the polarization of stars to trace the same $B$-field as the $\phi_{\rm synch}$ (see Section \ref{sec:results_loopIV}). 

Finally, region $\rm IV_A$ shows alignment for the bins centered on [195, 365]\,pc. As discussed further in Section\,\ref{sec:results_loopIV}, the alignment between stellar polarization and polarized dust emission also occurs at comparable distances. We can thus place an upper limit on the location of the synchrotron emission at $ \leq$195\,pc. The condition for the lower distance limit is found at 165\,pc. Our estimate of the location of the region's synchrotron emission is [165, 195]\,pc.

The distance limits for all regions are summarized in Table \ref{tab:regions}. For ease of use we report the middle of each limiting distance range as the distance to the feature with uncertainties spanning the full range of the distance bracket.

\section{Discussion}
\label{sec:discussion}

We have compared the polarization angles of synchrotron and dust emission with measurements of starlight polarization as a function of distance to obtain limits on the location of two synchrotron features: Loop\,I and Loop\,IV, as defined by \cite{Vidal2015}. We now discuss our findings for the distance to these structures in the context of the literature for each loop.

\subsection{Loop\,I} 

The distance to Loop\,I and its brightest part (the NPS) has been the subject of numerous studies \citep[for recent literature reviews see][]{Planck2015_XXV,Dickinson2018,Kataoka2018,Shchekinov2018}.

The high Galactic latitude part of Loop\,I ($b>40^\circ$) was first measured to be at a distance of $100\pm20$\,pc, on the basis of comparing the optical polarization of $\sim 10$ stars with synchrotron polarization \citep{Bingham1967}. Subsequently, \citet{Spoelstra1972} confirmed that the observed correspondence between optical and radio polarization holds for stars in the distance range 50--100 \,pc. \citet{ellisaxon1978} noted the correspondence remains for stars in the range 50--300\,pc. As more optical measurements became available, the correlation has been noted in multiple subsequent works \citep[e.g.,][]{leroy1999,Santos2011,Berdyugin2014}. Our results for regions I$_{\textrm{B}}$, I$_{\textrm{C}}$, and I$_{\textrm{D}}$, are in excellent agreement with these previous determinations. We do note, however, that the aforementioned literature estimates were based on the comparison of individual stars with synchrotron polarization angles and as such did not budget for uncertainties due to beam effects or intrinsic scatter of stellar measurements within the selected sky regions. We have shown in Sections \ref{sec:results_loopI} and \ref{sec:limits} that both factors come into play when considering the angle differences $\theta^*-\phi_{\rm synch}$. The distance limits that we have placed are conservative and account for these uncertainties. More precise determinations would require a better sampling of stellar measurements both in distance and on the plane of the sky.

Through morphological comparison, the NPS has been associated with a region of bright X-ray emission extending from the Galactic plane to high Galactic latitude \citep[e.g.,][]{BorkenIwan1977,Egger1995,Snowden1995}. Other determinations of the distance to Loop\,I rely on morphological correlation of either the radio Loop\,I and/or the X-ray NPS with different tracers, including \ion{H}{1} line emission \citep[e.g.,][]{Berkhuijsen1971}, stellar extinction \citep[e.g.,][]{Reis2008}, as well as dust emission and FUV lines \citep[e.g., see maps in ][]{Park2007}. \ion{H}{1} structures in the vicinity of the NPS at latitudes $b>55^\circ$ are located at 95--157\,pc, shown by interstellar absorption lines in stellar spectra \citep{PuspitariniLallement2012}. Stellar extinction towards the high latitude dust neighboring the NPS ($b = 26^{\circ}$--$55^{\circ}$) arises in the range 70--135\,pc \citep{Das2020}. These determinations are extremely precise \citep[e.g.,][give systematic uncertainties of 4--7\,pc towards individual regions]{Das2020}. Using existing stellar polarization data alone, the distance determination is an order of magnitude less precise. Despite this shortcoming, there is a significant added benefit in using stellar polarization, as we argue next.

At high latitude, the stellar polarization data tie together the dust-bearing, cold medium that forms \ion{H}{1} arches and causes the extinction and polarization to nearby stars \citep{Iwan1980,PuspitariniLallement2012,Das2020} with the radio Loop\,I through the excellent observed correlation between the stellar and radio polarization angles. A coincidence between the spatial distribution of the cold gas with the radio emission is much harder to argue for \citep[e.g., as done in][]{Lallement2018} if one takes into account this long-standing observational fact. The alignment between stellar and synchrotron polarization angles starting at $\sim$100\,pc (Figs.\,\ref{fig:loopI_lower}, \ref{fig:appendix}) in regions I$_{\textrm{B}}$ and I$_{\textrm{C}}$ ($60^{\circ} > b > 30^\circ$) leaves little room for contribution of background emission to the polarization at 30\,GHz. At the highest latitude range studied (region I$_{\textrm{D}}$) more stellar polarization data are needed to reduce observational uncertainties and confirm whether this conclusion is valid there as well. 

From the above reasoning, radio synchrotron polarization, \ion{H}{1} emission, dust emission/stellar absorption and optical polarization data are definitely linked. For the X-ray data, however, the link is less strong. Whether the X-ray NPS at high latitude is physically connected to the radio Loop\,I and (therefore to the above tracers) can be argued mainly on the basis of morphological similarity \citep[e.g.,][and references therein]{Iwan1980}. 
Another potential means for connecting X-rays with cold gas/dust tracers is the column density inferred by modeling the X-ray spectrum, which has been used to place tight constraints for the distance to the structure at low latitudes \citep[$b < 9^\circ$, ][]{Lallement2016,Das2020}. \citet{Akita2018} modeled \textit{Suzaku} data of the NPS at high latitude, finding that the necessary X-ray absorbing column is consistent with the total column density along the line of sight. Their finding is not particularly constraining, as the total column is contained within 75\,pc at these latitudes, as shown using stellar extinction data by \cite{Das2020}. Consequently, morphological similarity is the only argument that remains to suggest that the high-latitude X-ray NPS and the radio Loop\,I are parts of the same structure -- a point we will return to shortly.

The finding that at high latitude the polarized radio Loop\,I is at $\sim$100\,pc is in apparent contrast with the conclusions drawn from recent studies that analysed X-ray data \citep[][]{Akita2018,LaRocca2020} and Faraday rotation \citep[][]{Sun2015,XuHan2019}. There are three arguments put forward to argue that the NPS is further than 200\,pc.

The first argument, put forward by \citet{XuHan2019}, relies on the dispersion measure of the pulsar PSR
J1503+2111 $(l,b) = (29^\circ\!.1,59^{\circ}\!.3$). These authors derive a mean electron density towards the pulsar of 0.013\,cm$^{-3}$, by assuming a uniform ISM density. This value is then compared to a model for the electron density of Loop\,I presented in \citet{Yao2017}. Because the inferred electron density towards the pulsar is found to be two orders of magnitude less than the model, the authors conclude that Loop\,I must be background to the pulsar. However, the ISM density is not uniform along the 240\,pc sightline towards the high-latitude parts of Loop\,I. In fact, the dust distribution shows an abrupt rise at about 100\,pc, from both starlight polarization and extinction \citep[Fig.\,\ref{fig:loopI_lower} and also][]{Das2020}. We recalculate the electron density here. We know that almost 100\% of the dust column is within $\sim$70\,pc \citep[highest latitude region in table\,2 of][]{Das2020}, most of it localized in a thin shell\footnote{This agrees with the 3D dust map of \citet{Leike2020} for all regions except $\rm I_D$, for which we see a gradual rise in extinction in Fig.\,\ref{fig:loopI_lower}. The extinction values in this region are very low and could be prior-dominated.}. Taking the thickness to be 1--10\,pc, from the DM we would infer an electron volume density of 0.3--3\,cm$^{-3}$, which is at most a factor of 3 different to the \cite{Yao2017} model (1.9\,cm$^{-3}$). In addition to this, the comparison to the \citet{Yao2017} model may be erroneous, since the data show negligible dispersion measure at high latitude, with most pulsars constraining the model being at latitude $b<20^\circ$. We conclude that the observed DM towards this pulsar is not in tension with the stellar polarization determination of the distance to the high-latitude parts of Loop\,I.

The second argument, presented in \cite{Akita2018}, uses X-ray spectral information. The authors assume a uniform density throughout the line-of-sight and a plausible volume density of electrons for the X-ray emitting gas in the NPS. One can then divide the measured emission measure from X-ray data by the square of the assumed volume density to infer the line-of-sight path length. This leads to an estimate of 1--8\,kpc, depending on the assumed volume density, which the authors use to argue in favor of a Galactic center distance to the NPS. However, for the same reasons discussed above, a uniform density assumption for the medium towards the NPS is unjustified. The fact that the emission measure depends on the square of the electron volume density, means that any clumpiness (i.e., volume filling factor less than unity) would accentuate the problem. Such assumptions can lead to erroneous conclusions, in contrast to the high-confidence measurements  from stellar data. 

The third argument was made by \citet{Sun2015}, based on Faraday rotation of the diffuse synchrotron emission. These authors find that the Faraday depth towards the NPS at high latitude is consistent with zero, with all of the observed Faraday depth being attributed to the medium in the background of the NPS. Since the $B$-field is expected to lie parallel to the Galactic plane, they conclude that the $B$-field at high latitude will have only a small line-of-sight component, causing the Faraday depth to rise slowly as a function of distance. Thus, they expect the NPS at $b > 50^\circ$ to be several hundred parsecs away. The behavior of the Faraday depth at lower latitude is very different, which leads them to propose that the emission from lower latitudes arises from a longer path length.

As a final note, we mention one more argument that has been presented against the local SNR model for Loop\,I. \cite{Sofue2000} argue that if the structure is local, it would be the only one of its kind because its physical diameter of $\sim 200$\,pc would be much larger than that of typical SNRs. We note that structures of similar size exist in the local ISM: a prominent example is the Orion-Eridanus superbubble, at a distance of 400\,pc. It has a diameter of 200\,pc and was created by the combination of energy output from multiple young stars and possibly supernova explosions \citep[see][and references therein]{Soler2018}. Similar activity from the Sco-Cen association could easily have created this very large nearby bubble that is the North Polar Spur \citep[e.g., as suggested by][]{Egger1995}, meaning that Loop\,I may only be one of many reheated SNRs. 

From the above discussion, we conclude that there remains no contradiction between the measurements of the distance to Loop\,I at high latitude and the evidence presented in the literature -- the contradiction is in the conclusions that have been drawn. At lower latitude ($b \lesssim 10^{\circ}$), the situation is even more complicated, with many works using X-ray data \citep{Akita2018,Kataoka2021,Predehl2020} and concluding that the NPS is located near the Galactic center, in support of the model proposed by \cite{Sofue1977,Sofue2000}. These works assume that Loop\,I (and similarly the NPS which is associated with it) arise from a single structure. However, as argued by \citet{Dickinson2018}, if this assumption is lifted, then the superposition of multiple structures can explain the low latitude parts of Loop\,I, without the need to ignore/refute the very accurate nearby distance of the high latitude portion. Indeed, evidence has accumulated that this line-of-sight confusion is prevalent at low latitudes.

Several lines of evidence point to a superposition of multiple structures along the line of sight that gives rise to the radio and X-ray emission at intermediate to low latitude. For example, at latitudes $b < 9^\circ$, \cite{Lallement2016} find that the X-ray data require an absorbing column that places the emission further than 800\,pc. Indeed, \citet{Lallement2018,Das2020} show the existence of multiple dust structures along the line-of-sight towards these low latitude regions. \citet{Sofue2016} compare the morphology of X-ray data with the radio continuum and conclude that the X-rays are absorbed by the Aquila Rift clouds (below $b < 30^\circ$), and therefore must be beyond 1\,kpc. Our results for region I$_{\textrm{A}}$, at latitude $10^{\circ}$--25$^{\circ}$ show the lack of alignment between stars and synchrotron polarization angles, and are consistent with these conclusions. 

While we are not able to disentagle the effect of the NPS at low latitude from background contributions, better sampled data might be able to do so. For example, \citet{Jones2016} show that NIR stellar polarization is perpendicular to the Galactic Plane at the base of the NPS ($b \sim 0.2^\circ$) at distances of a couple hundred parsecs.

\begin{figure*}[ht]
    \includegraphics[scale=1]{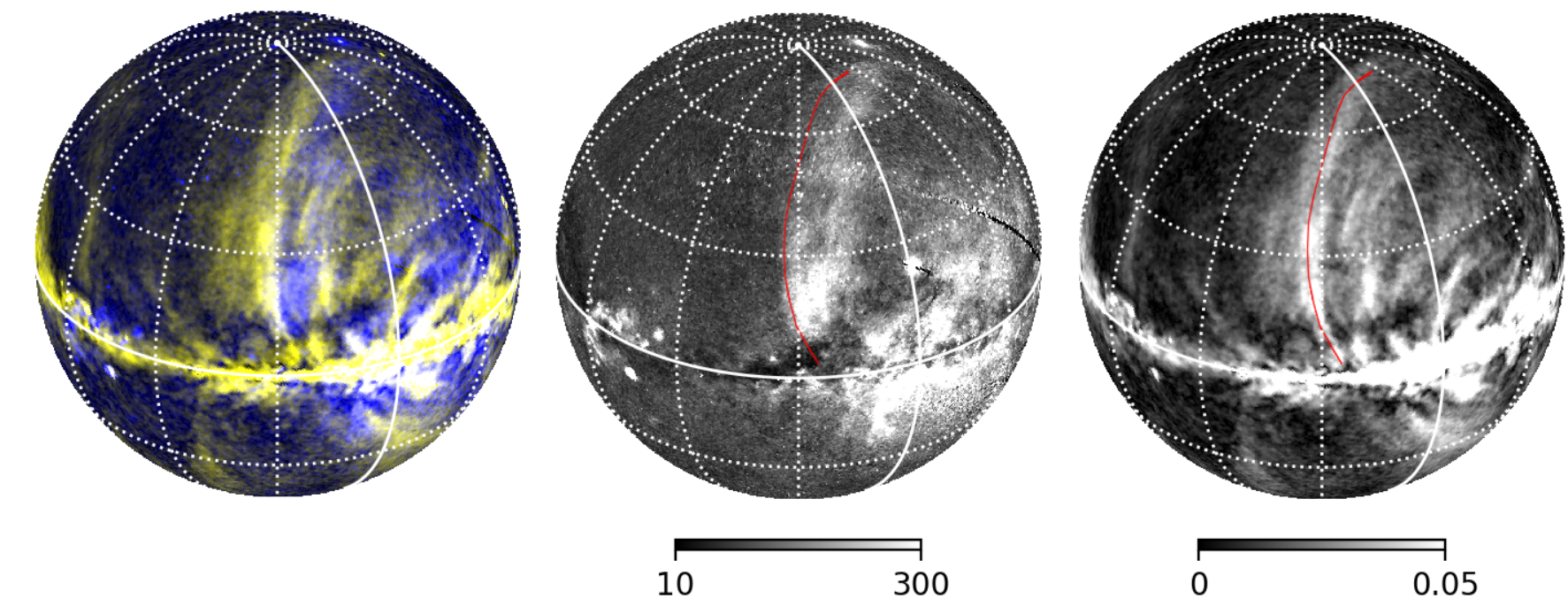}
    \caption{Comparison between radio and X-ray data. Left: Composite image with synchrotron polarized intensity (this work, yellow), and X-rays from ROSAT's R6 0.73–1.56 keV band \citep{Snowden1997} (blue). Middle: X-ray map of the ROSAT R6 band (units of $10^{-6} \rm counts/s/arcmin^2$) with the outline of radio Loop\,I from \citet{Vidal2015} marked with a red line. Right: Polarized intensity of synchrotron emission (this work) with the outline of  Loop\,I as in the middle panel (units of mK). All images have a linear colorscale, $N_{\rm side} = 128$, and are in Orthographic projection centered at $(l, b)\, =\, (30^\circ, 30^\circ)$. Grid lines are spaced by $30^\circ$. Lines of constant galactic longitude, $l = 0^\circ$ and constant latitude, $b = 0^\circ$, are marked solid white. }
    \label{fig:rgb}
\end{figure*}

In Fig.\,\ref{fig:rgb} we examine the morphologies of the polarized radio intensity and X-ray data from ROSAT (band B6 at 0.73--1.56\,keV). The well-known offset between radio Loop\,I and the X-ray NPS \citep[e.g.,][]{Kataoka2021} is best seen by comparing the relative location of the emission to the line that traces the peak intensity of the polarized emission of Loop\,I from \citet{Vidal2015}. This line neatly bounds the bright NPS as seen in the X-ray map. This remarkable morphological similarity, strongly suggests that the X-rays and radio are part of the same object, as previously noted in various works \citep[e.g.,][]{BorkenIwan1977}. This would mean that our distance constraints on the Loop\,I polarized radio emission apply to the X-ray NPS as well (for the regions studied, namely $b > 30^\circ$).

We propose the following hypothesis for explaining the combination of radio, optical and X-ray data, where the apparent contradiction between distance estimates is more prominent in the recent literature. At high latitude, the bulk of the radio polarization, the extinction, and stellar polarization are all local (within $\approx 100$\,pc), as necessitated by the measurements of \citet{PuspitariniLallement2012,Das2020} and this work.
However, not all of the high-latitude X-rays attributed to the NPS need be local. The recent e-ROSITA finding of a symmetric, but much fainter X-ray lobe in the South \citep{Predehl2020} leads us to the hypothesis that \textit{there is an equally faint Northern Galactic center lobe, on top of which there is superimposed local X-ray emission.} As argued by several authors \citep[e.g.,][]{BorkenIwan1977,Iwan1980}, the radio and X-ray emission requires a reheating episode of an old SNR by a more recent event. The asymmetric appearance of the radio Loop\,I is easily explained by the local reheated SNR hypothesis \citep[e.g.,][]{Heiles1980,Wolleben2007}. If the entire X-ray emission were due to activity in the Galactic center\footnote{The X-ray e-ROSITA emission has been suggested to be associated with the Fermi Bubbles  \citep[][]{Predehl2020}.}, the observed asymmetry between the two hemispheres would require a corresponding asymmetry in the Galactic halo \citep[the argument proposed by][]{Sarkar2019}. In our view, the local SNR hypothesis for the high-latitude part of the NPS is not only a more \textit{plausible} explanation, it is the only explanation that is consistent with the wealth of observational constraints on the distance to the structure, as discussed above.

\subsection{Loop\,IV} 
In contrast to Loop\,I, the distance to Loop\,IV has received much less attention. \cite{ellisaxon1978} searched for a signature of Loop\,IV in the polarization of stars but were unsuccessful due to lack of stellar  measurements. With improved statistics we are able to revisit the relation between the radio emission and starlight polarization in parts of Loop\,IV. Our analysis has relied critically on the comparison between \textit{all three} tracers of the magnetic field: starlight, synchrotron and dust emission polarization angles. 

We find evidence that Loop\,IV is likely a superposition of discrete structures along the line-of-sight. To the best of our knowledge, this is the first evidence to date against the common assumption of Loop\,IV being a spherical shell, visible as a small circle in the sky. In hindsight, another indication of this can be traced back to \cite{HeilesJenkins1976}, who examined the morphology of \ion{H}{1} gas towards Loop\,IV. They found some \ion{H}{1} structures that morphologically resembled parts of Loop\,IV, but there were no features that matched the radio continuum morphology over the entire circumference of Loop\,IV.

\citet{Spoelstra1973} estimated a distance to radio Loop\,IV by fitting a spherical SNR model \citep[]{vanderLaan1962} to the total intensity at 1.4\,GHz. He found a value of 250\,pc for the distance to the center of the sphere. Given the observed radius in the sky of $20^{\circ}$, this would place the southern tangential endcap of the sphere (our regions IV$_{\textrm{A}}$ and IV$_{\textrm{B}}$) at a distance of 235\,pc. This is not consistent with our distance limits on region $\rm IV_A$ of [165, 195]\,pc.
The results of \citet{Spoelstra1973} rely on the model assumptions of a spherical shell expanding in a homogeneous medium. As discussed in Section\,\ref{sec:results_loopIV}, our multi-tracer comparison of polarization angles strongly suggests that Loop\,IV is not a single spherical structure.

Based on the Loop\,IV distance estimate of \citet{Spoelstra1973}, \citet{BorkenIwan1977} and \citet{Iwan1980} suggested that Loop\,IV could be a young SNR that is interacting with Loop\,I (assumed to be a Myr-old SNR) and re-heating it. Our distance limits on Loop\,IV would place part of it (region $\rm IV_A$) at a larger distance of [165, 195]\,pc compared to the distance of [95, 130]\,pc (our tightest limit on the distance to Loop\,I). With increased stellar polarization measurements we will be able to improve upon these constraints. We note that a recent analysis of $\gamma$ rays also indicates a nearby distance to both Loop\,I and IV \citep{Johanesson2021}.

\section{Conclusions}
\label{sec:conclusions}

We have combined optical stellar polarization with polarized dust and synchrotron emission to investigate the correlation of these magnetic field tracers towards high-latitude synchrotron spurs. Our findings are listed below:

\begin{itemize}
    \item There is an overall tight correlation between the polarization angles of synchrotron, dust emission and stars at high galactic latitude, confirming earlier comparisons.
    
    \item In regions where the tracers $\theta^*$, $\phi_{\rm synch}$ and $\chi_{\rm dust}$ are well-aligned, we have used their correlation as function of stellar distance to place distance limits on Loop\,I and Loop\,IV.
    
    \item For Loop\,I we find that $\theta^*$ and $\phi_{\rm synch}$ become well aligned at a distance of $\sim100$\,pc, confirming earlier evidence that the high-latitude portion of this structure is nearby. At the lowest latitude portion of the Loop studied here ($b \sim 20^\circ$) $\theta^*$ and $\phi_{\rm synch}$ are not statistically aligned and their relative orientation varies with distance, suggesting the presence of overlapping features along the line of sight (Sections \ref{sec:results_loopI} and \ref{sec:limits}).
    
    \item For Loop IV we find a tight correlation between $\chi_{\rm dust}$ and $\phi_{\rm synch}$ for only part of the structure, for which we place distance limits. The observed lack of correlation towards a significant part of the loop suggests that radio Loop\, IV results from a superposition of unrelated  structures along the line of sight (Sections \ref{sec:results_loopIV} and \ref{sec:limits}).
    
    \item We propose a hypothesis that can reconcile the apparent inconsistency between the distance to the NPS from X-ray data and other distance constraints (Section \ref{sec:discussion}).
\end{itemize}

The main limitation in our analysis is the lack of starlight polarization coverage. Further progress on placing distance limits to the majority of synchrotron loops will require an order of magnitude increase of stellar measurements in these regions, in order to obtain adequate statistics with distance. New data are anticipated over the next few years, which will open up large-scale multi-tracer studies of the Galactic magnetic field and structure. Dedicated optical starlight polarization surveys will improve the depth, accuracy, and most importantly, the sky coverage. The PASIPHAE survey \citep[e.g.,][]{pasiphae} aims to measure the polarization of millions of stars towards high Galactic latitude, producing more than 100 stellar measurements per square degree. Higher signal-to-noise ratio all-sky synchrotron polarization maps will also soon be available at 5\,GHz from the C-Band All-Sky Survey \citep[C-BASS;][]{Jones2018}, which will provide almost full-sky coverage with high S/N and minimal contamination from Faraday Rotation. Combining these maps with existing data from \textit{Planck}, WMAP, the S-band Polarization All Sky Survey \citep[at 2.3\,GHz,][]{Carretti2019} and the Q-U-I Joint Tenerife Experiment \citep[at 10--42\,GHz,][]{Rubino-Martin2012b} will allow for much more precise determinations of synchrotron polarization angles at high latitude. At even lower frequencies ($\sim 100$\,MHz), data from the Low Frequency Array (LOFAR) will enable a tomographic view of the synchrotron emission, opening up new avenues to disentangle the contribution from multiple emitting components along the line of sight \citep{vanEck2019}.

\acknowledgments
We thank Samir Johnson for contributing to discussions in the early stages of this work. We thank Matias Vidal for providing coordinates to spurs and relevant code. We thank Juan Soler for providing helpful suggestions as well as the anonymous reviewer for a very thorough reading of our manuscript.
Support for this work was provided by NASA through the NASA Hubble Fellowship grant  \#HST-HF2-51444.001-A  awarded  by  the  Space Telescope Science  Institute,  which  is  operated  by  the Association of Universities for Research in Astronomy, Incorporated, under NASA contract NAS5-26555. The work at the California Institute of Technology was supported by the National Science Foundation (NSF) awards AST-0607857, AST-1010024, AST-1212217, and AST-1616227, AST-1611547, and by NASA award NNX15AF06G.  
CD acknowledges support from an STFC Consolidated Grant (ST/P000649/1). CD would like to thank Caltech for their hospitality during extended visits.

Based on observations obtained with Planck, an ESA science mission with instruments and contributions directly funded by ESA Member States, NASA, and Canada.
This work has made use of data from the European Space Agency (ESA) mission
{\it Gaia}, processed by the {\it Gaia}
Data Processing and Analysis Consortium (DPAC). Funding for the DPAC has been provided by national institutions, in particular the institutions participating in the {\it Gaia} Multilateral Agreement.

\vspace{5mm}
\facilities{WMAP, {\it Planck}, {\it Gaia}}

\software{astropy \citep{astropy:2018},  
          HEALPix \citep{Gorski2005},
          healpy \citep{Zonca2019}
          }

\pagebreak

\bibliography{bibliography}{}
\bibliographystyle{aasjournal}


\end{document}